\documentclass[%
aps,
superscriptaddress,onecolumn,
 amsmath,amssymb,
 aps,
]{revtex4}

\usepackage{pdfpages}
\usepackage{subfigure}
\usepackage{bm}
\usepackage{graphicx}
\usepackage{hhline}

\begin{document}


\title[Cold atom interferometer for space gravity gradiometry]{Concept study and preliminary design of a cold atom interferometer for space gravity gradiometry}

\author{A. Trimeche}
\affiliation{LNE-SYRTE, Observatoire de Paris, Universit\'e PSL, CNRS, Sorbonne Universit\'e, \\61 avenue de l'Observatoire 75014 Paris, France}

\author{B. Battelier}
\affiliation{LP2N, Laboratoire Photonique, Num{\'e}rique et Nanosciences, Univ. Bordeaux -- IOGS -- CNRS:UMR 5298, F-33400 Talence, France}

\author{D. Becker}
\affiliation{Leibniz Universit\"at Hannover, Institut f\"ur Quantenoptik, Welfengarten 1, 30167 Hannover, Germany}

\author{A. Bertoldi} 
\author{P. Bouyer}
\affiliation{LP2N, Laboratoire Photonique, Num{\'e}rique et Nanosciences, Univ. Bordeaux -- IOGS -- CNRS:UMR 5298, F-33400 Talence, France}

\author{C. Braxmaier}
\affiliation{ZARM, University of Bremen, Am Fallturm 2, 28359 Bremen, Germany}
\affiliation{DLR Institute for Space Systems, Robert-Hooke-Strasse 7, 28359 Bremen, Germany}

\author{E. Charron}
\affiliation{ISMO, CNRS, Univ. Paris-Sud, Universit\'e Paris-Saclay, F-91405, Orsay cedex, France}

\author{R. Corgier}
\affiliation{Leibniz Universit\"at Hannover, Institut f\"ur Quantenoptik, Welfengarten 1, 30167 Hannover, Germany}
\affiliation{ISMO, CNRS, Univ. Paris-Sud, Universit\'e Paris-Saclay, F-91405, Orsay cedex, France}

\author{M. Cornelius}
\affiliation{ZARM, University of Bremen, Am Fallturm 2, 28359 Bremen, Germany}

\author{K. Douch}
\thanks{Present address: Geod{\"a}tisches Institut Stuttgart (GIS), Universit{\"a}t Stuttgart, Geschwister-Scholl-Str. 24/D, 70174 Stuttgart, Germany}
\affiliation{Institut f{\"u}r Erdmessung (IfE), Leibniz Universit{\"a}t Hannover, Schneiderberg 50, 30167 Hannover, Germany}

\author{N. Gaaloul}
\affiliation{Leibniz Universit\"at Hannover, Institut f\"ur Quantenoptik, Welfengarten 1, 30167 Hannover, Germany}

\author{S. Herrmann}
\affiliation{ZARM, University of Bremen, Am Fallturm 2, 28359 Bremen, Germany}

\author{J. M{\"u}ller}
\affiliation{Institut f{\"u}r Erdmessung (IfE), Leibniz Universit{\"a}t Hannover, Schneiderberg 50, 30167 Hannover, Germany}

\author{E. Rasel}
\author{C. Schubert}
\affiliation{Leibniz Universit\"at Hannover, Institut f\"ur Quantenoptik, Welfengarten 1, 30167 Hannover, Germany}

\author{H. Wu}
\affiliation{Institut f{\"u}r Erdmessung (IfE), Leibniz Universit{\"a}t Hannover, Schneiderberg 50, 30167 Hannover, Germany}

\author{F. Pereira dos Santos}
\affiliation{LNE-SYRTE, Observatoire de Paris, Universit\'e PSL, CNRS, Sorbonne Universit\'e, \\61 avenue de l'Observatoire 75014 Paris, France}


\begin{abstract}

We study a space-based gravity gradiometer based on cold atom interferometry and its potential for the Earth's gravitational field mapping. The instrument architecture has been proposed in [Carraz et al., Microgravity Science and Technology \textbf{26}, 139 (2014)] and enables high-sensitivity measurements of gravity gradients by using atom interferometers in a differential accelerometer configuration. We present the design of the instrument including its subsystems and analyze the mission scenario, for which we derive the expected instrument performances, the requirements on the sensor and its key subsystems, and the expected impact on the recovery of the Earth gravity field.

\end{abstract}



\maketitle

\section{Introduction}

Satellite gravimetry missions, such as the GRACE (Gravity Recovery And Climate Experiment) \cite{Tapley2004} and the GOCE (Gravity field and steady-state Ocean Circulation Explorer) \cite{Drinkwater2003, Johannessen2003} missions, have revolutionized our knowledge of the gravity field over the whole Earth surface and our understanding of mass redistribution and mass transport processes on a global scale. In particular, the GOCE mission, launched in 2009 and active up to 2013, carried a gravity gradiometer on-board a satellite for the first time. It allowed for a precise measurement of the static gravity field with unprecedented accuracy and spatial resolution. The geoid was determined with an accuracy of about 1 to 2 cm for a spatial resolution of 100 km on the Earth surface \cite{Brockmann2014a}. By providing the Earth gravity field down to small spatial scales, our understanding of a number of processes related to solid-earth physics and ocean circulation has been greatly improved \cite{Knudsen2011} and the global unification of height systems \cite{Rummel2014} could be implemented. GOCE also brought new and unexpected scientific results in seismology, space weather and changes in ice masses. In this mission, differential accelerations measured on-board a single satellite with an ensemble of ultra-sensitive electrostatic accelerometers allowed to determine all components of the gravity gradient tensor, with best sensitivities in the range of 10-20 mE/Hz$^{1/2}$ in the measurement bandwidth (i.e. 5-100 mHz), out of which models of the gravity field could be reconstructed.


In this article, we show that the use of cold-atom-based gravity gradiometers, on-board a dedicated satellite at low altitude (250-300 km), can meet the requirements to improve our present knowledge of the Earth gravity field. In particular, the GOCE gravity gradients showed poor performance in the lower frequency band, where the noise power spectral density (PSD) increases with the inverse of the frequency. Dealing with this low-frequency noise is a great challenge for gravity field recovery, where special decorrelation filters were tailored and used for whitening the noise \cite{Schuh2003}. This will not be a problem for a  gravity gradiometer based on a cold atom interferometer (CAI), as it naturally provides gravity gradients with white noise at all frequencies, except for very high frequencies (i.e. above 100 mHz) which is not relevant for gravity field recovery from space. Moreover, this novel atom-based gradiometer is expected to provide gravity gradients with an improved sensitivity level of the order of 5 mE/Hz$^{1/2}$.

The article is structured as follows. We start in section \ref{sec:principle} by describing the instrument and measurement principle, which relies on a state-of-the-art manipulation sequence of the atomic source described in \ref{sec:source}. We then calculate in section \ref{sec:phase} the phase shift of the interferometer for arbitrary gradients and rotation rates, in the simplified case of a circular orbit around a spherical and homogeneous Earth. This allows to derive in section \ref{sec:error} specifications for the control of the atomic source parameters and for the attitude control of the satellite. A Monte Carlo simulation of the interferometer is then presented in section \ref{sec:model}, which allows us for accounting in a comprehensive way for the geometry of the interferometer and furthermore to evaluate the loss in sensitivity as well as the amplitude of several systematic effects due for example to the finite size of the laser beams and the atomic cloud or the finite duration of the interferometer pulses. The results of this simulation are used to refine key specifications for the laser system setup and the atomic source, in order to keep parasitic differential phase shifts (both noise and systematic) below the target uncertainty. 

Section \ref{Sec:Design} is dedicated to the instrument design and related engineering aspects. Details on the design of critical elements and subsystems are given, in particular on the retroreflecting mirror, on the laser, vacuum and detection systems. Engineering tables are elaborated. Finally, we evaluate in section \ref{Performance_analysis} the impact of the sensor performance for gravity field recovery. This is performed thanks to numerical simulations of the measured gravity in the presence of realistic noise for the sensor and the control of the attitude of the satellite.


\section{Principle of the measurement}
\label{sec:principle}
The concept of the gradiometer is based on the geometry proposed in \cite{Carraz2014}, which measures differential acceleration with two spatially separated atom interferometers (AIs) \cite{Snadden1998}. The interferometers are realized using a sequence of three light pulses based on stimulated Raman transitions. The momentum transfer provided by the Raman diffraction process allows the splitting, redirection and recombination of the atomic wave packets along two paths, thus creating an atomic analogy of a Mach-Zehnder interferometer. In such atom interferometers~\cite{Berg2015PRL,Louchet2011NJP,Gauguet2009PRA,Kasevich91PRL}, the atomic populations in the two output ports are modulated with the phase difference accumulated along the two paths. For an acceleration $a$ along the direction of the laser beams, this phase is given by $\Phi=kaT^2$, where $\hbar k$ is the momentum imparted by the Raman transitions onto the atoms, and $T$ is the free evolution time in between two consecutive Raman pulses. Performing differential acceleration measurements with two such interferometers separated by a distance $D$ allows extracting the gravity gradient $\gamma$ out of the differential phase shift $\Delta\Phi=\Phi_2-\Phi_1=k(a_2-a_1)T^2=k\gamma D T^2$, with $a_1$ and $a_2$ the accelerations experienced by the atoms in the two interferometers~\cite{McGuirk02PRA}. Moreover, using the same Raman lasers for the two interferometers enables a high rejection ratio to common mode sources of noise and systematic effects~\cite{McGuirk02PRA,Bertoldi2006}. More details on the working principle of the interferometers will be given in section \ref{doublediffraction}. 

Assuming an interferometer phase noise at the mrad/shot level, the corresponding sensitivity to the gradient is in the mE/shot range, for a pulse separation time $T=5$~s and a distance $D=0.5$~m. To take a full advantage of this excellent single-shot sensitivity, a high measurement rate is desirable, which can be achieved by interrogating several atomic clouds at the same time \cite{Biedermann2013,Dutta2016}. This interleaved scheme requires to produce atomic sources with a cycle time significantly shorter than the interferometer duration \cite{savoie2018interleaved}. With a production time of about 1~s, the corresponding sensitivity would lie in the low mE/Hz$^{1/2}$, which compares favourably with the ultra sensitive electrostatic gradiometers of the GOCE mission. 

\begin{figure}[!ht] 
\begin{center}
\includegraphics[width=0.9\textwidth]{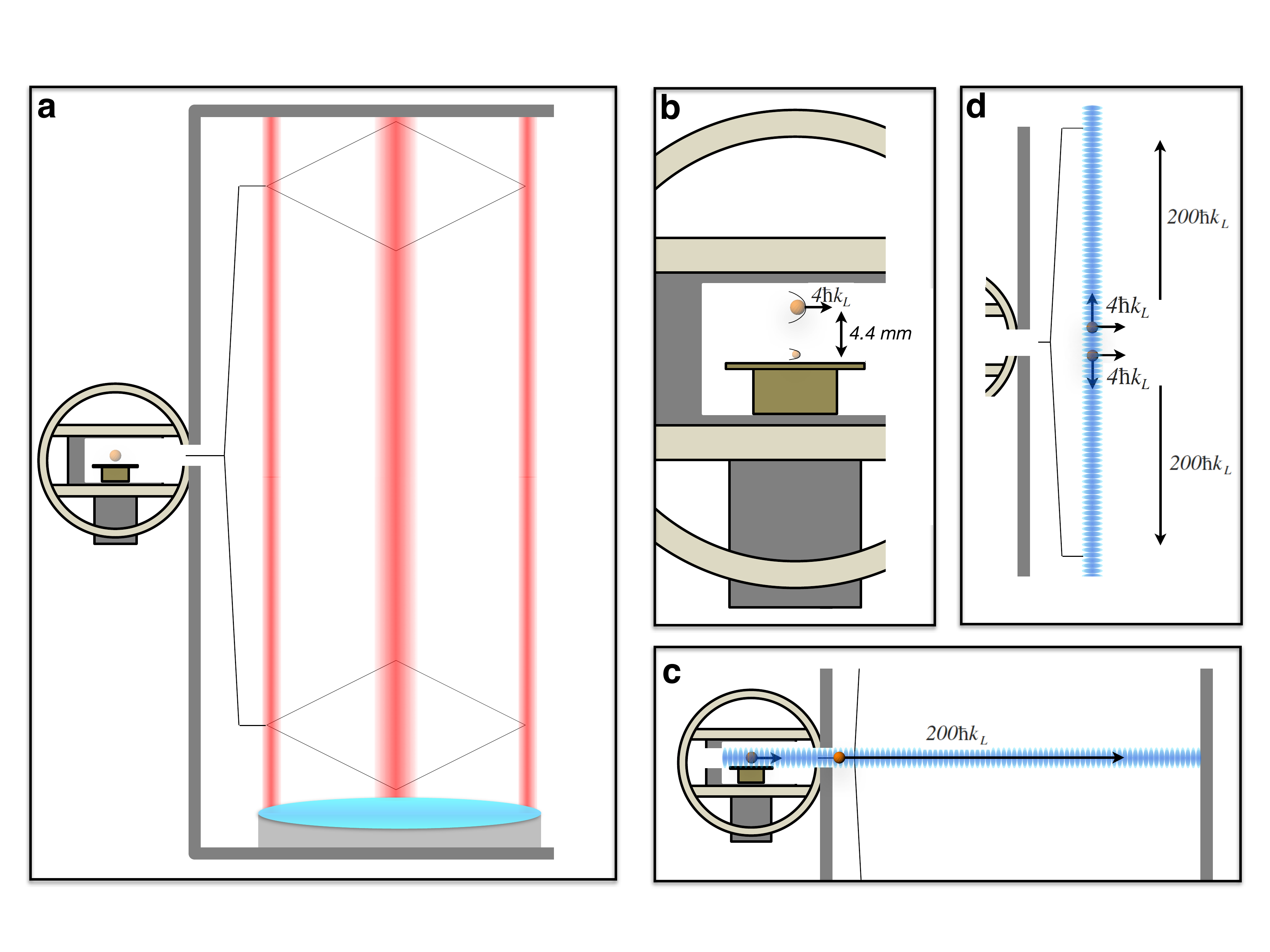}
\caption{(a) Scheme of the gravity gradiometer, based on differential accelerometry with two separated atom interferometers. (b) An initial BEC source of $10^6$~atoms is magnetically evaporated, displaced and collimated in $1.1$~s. (c) Horizontal transport step to the  interferometry chamber (12\,cm in 100\,ms). (d) The BEC is split in two by the combination of a double Raman diffraction and a twin-lattice technique feeding both interferometers with ensembles at a horizontal velocity of 4 recoils.}
\label{Concept-Scheme}
\end{center}
\end{figure}

\section{Source preparation}
\label{sec:source}
The presented measurement principle is not limited to a specific atomic species. It requires, however, an atomic source production at a high flux ($10^6$~atoms/s) to reach the targeted phase noise of 1\,mrad per cycle of 1\,s, yet at a low expansion rate (0.1~mm/s) characteristic of near-condensed regimes. We analyzed possible candidates of high-flux degenerate sources in terms of their technical readiness for the proposed space-borne gravity gradiometer. Sources based on alkaline atoms (e.g. Li, Na, K, Rb, Cs) are widely used in cold-atom experiments and have shown excellent performance. Due to their rather simple energy levels structure, there are several applicable laser cooling schemes applicable and evaporative cooling to degeneracy is possible both in magnetic and optical potentials~\cite{Ketterle1995,Mewes-PRL-1997,Clement2009,Naik2018}. In particular, Rubidium sources have been established as reliable sources for atom interferometry fundamental physics experiments as well as geodetic applications~\cite{zhou2015test,rosi2014precision,hauth2013first}. High-flux evaporative cooling of Rb ($>10^5$~atoms/s) has been shown both with atom chips~\cite{Rudolph2015} or dipole traps~\cite{PhysRevA.71.011602} and atom interferometry can be performed with either Raman or Bragg transitions in single or double diffraction configurations~\cite{le2008limits,Leveque2009,altin2013precision,ahlers2016double}. Recently, alkaline-earth-like atoms (e.g. Sr, Yb) have been successfully cooled to degeneracy~\cite{stellmer2013laser,fukuhara2007bose} and are considered to be promising candidates for high-precision interferometry as well. Thanks to their special energy levels structure, they are immune to residual quadratic Zeeman shifts, one of the dominant contributions to the uncertainty budget in alkaline-based atom interferometers. Moreover, interferometry can be performed on the clock transition to suppress technical laser phase noise thus increasing the performance of precision measurements e.g. weak equivalence principle test, gradiometry or gravitational wave detection~\cite{hogan2011atomic,Canuel2018}. The source flux of alkaline-earth-like atoms is, however, not yet at the same level of performance as the Rb ones. Besides of the performance of the source, the maturity of the required technology is crucial for a successful space mission ~\cite{Loriani2018GOAT}. The cooling and manipulation laser sources of Rb could be derived from two complementary systems, which are available and field-proven: compact diode lasers at 780~nm together with free-space optics~\cite{schkolnik2016compact} or fiber-based laser systems fed by frequency-doubled laser running at 1560~nm~\cite{theron2015narrow}. High-flux sources of condensed Rb have been already demonstrated in transportable and space qualified systems~\cite{Becker2018}. In addition, complexity, power consumption, size and mass considerations are in favor of a Rb-based choice compared to setups based on Sr or Yb. With recent progress made by the QUANTUS, MAIUS and CAL consortia ~\cite{abend2016atom,Becker2018,elliott2018nasa}, the atom chip solution has been assessed to be more advanced than the dipole trap one and will be the baseline for the proposed setup.

In order to span the required baseline of 50 cm together with a cycle time close to 1 s, we propose the following sequence for the transport and preparation of the atoms: we start by producing the  BEC (Bose-Einstein Condensate) atoms in close vicinity of the atom chip in a dedicated chamber (left frame a in figure~\ref{Concept-Scheme}). With an optimized atomic chip design, building on the work of \cite{Rudolph2015}, about $10^6$ ultracold atoms could be produced in less than 1~s (i.e. 800 ms) at a distance of few hundred microns. The created BEC is immediately magnetically displaced up to a distance of about 5 mm away from the chip surface (frame b in figure~\ref{Concept-Scheme}) using the external magnetic coils and shortcut-to-adiabaticity protocols as proposed in \cite{Corgier2018} and implemented in~\cite{rudolph2016matter}, allowing to reduce the duration of this transport down to about 200 ms. The geometry of the displaced trap can be tailored, thanks for instance to several layers of Z-shaped wires on the chip, to be almost spherically symmetric~\cite{li2018expansion}, with a final trapping frequency of about 15 Hz. These fast transports have the feature of inducing very low residual dipole oscillations in the final trap, which is an essential ingredient for the next steps. If the quantum gas needs, in addition, to be in its ground state, optimal control solutions are available and proved to be equally fast~\cite{amri2018optimal}. The atomic ensemble is released from this weak trap, freely expanding for 100\,ms before being collimated by a magnetic lens flashed for about 1.2\,ms~\cite{Muntinga2013}. This drastically reduces the expansion rate of the cloud, already at a Thomas-Fermi radius of $150~\mu$m, down to a calculated effective expansion energy of 100\,pK. The subsequent time evolution of the size of the BEC complies with the interferometric requirements on the atomic source. The control accuracy of the position and the velocity of the atomic clouds at the end of this first chip manipulation are estimated to be of the order of a fraction of a $\mu$m and $\mu$m/s, respectively
\cite{Corgier2018,rudolph2016matter,amri2018optimal}. 


To move the atomic ensembles into the interferometry region, a first Raman double diffraction initiates the cloud momentum at $4\,\hbar k$ as indicated in frame b of figure~\ref{Concept-Scheme}. This ensures a transport by a Bloch accelerated optical lattice (blue beam in figure~\ref{Concept-Scheme}(c)) that moves the atoms to the interferometry chamber by imparting, in few ms, a velocity corresponding to 200 recoils \cite{Cadoret2008}. Thanks to the atomic cloud collimation step, it is possible to restrict ourselves to the use of a small beam waist of 1-2 mm for the Bloch lattice and thus keep the power usage at the reasonable level of roughly 200 mW. Once the atoms reach the interferometry chamber (12\,cm in 100\,ms), the same optical lattice with opposite acceleration direction is used to decelerate the atoms to a final momentum of $4\,\hbar k$ as they started. 
At this point, the atom source has to be split into two halves to be moved upwards and downwards in order to feed the two interferometers as indicated in figure~ \ref{Concept-Scheme}(d). This is realized by combining the use of double-Raman-diffraction beams~\cite{Leveque2009} and twin-Bloch lattices \cite{abend2017chipgravi}. A first pair of retro-reflected Raman pulses splits the BEC to generate a pair of vertically moving momentum classes with $\pm 4 \hbar k$.  
By reflecting two Bloch lattices on the mirror, with 100 GHz relative detuning, one is able to create two running lattices similarly to what was done in~\cite{abend2017chipgravi}. The advantage of this highly symmetric scheme is that each of the initially Raman-split clouds will interact with the optical lattice moving in the same direction thanks to Doppler selectivity. The same treatment as in frame c) is subsequently pursued, with an acceleration up to 200 recoil velocities followed by a deceleration to a still momentum state. The transport distance needs, however, to be double the one of the horizontal step to reach 24 cm in 200 ms time.
In this manner, each of the gradiometer's Mach-Zehnder interferometers is fed by an incoming flux of atoms with  $4\,\hbar k$ initial velocity.

In total, the entire process of production, manipulation and transport of the BEC takes 1.4~s. However, a new magneto-optical trap (MOT) production can start as soon as the previous produced BEC has been loaded into the vertical lattices resulting in an effective cycle time of only 1.2~s. With that scheme it would be possible to run up to 8 interferometers simultaneously.

\section{Interferometer phase shift}
\label{sec:phase}
We calculate in this section the phase shift of a (single) atom interferometer linked to the frame of a satellite, orbiting at a fixed altitude with a constant orbiting frequency $\Omega_{\mathrm{orb}}$. The measurement axis is taken in the orbital plane, initially along $z$. We take into account the rotation rates of the satellite $\Omega_{x,y,z}$, which will allow to discuss the cases of nadir pointing, for which $\Omega_y = \Omega_{\mathrm{orb}}$ and $\Omega_{x,z} \simeq 0$, and inertial pointing, for which $\Omega_{x,y,z} \simeq 0$. The calculation of the atom trajectories is performed in the satellite frame, hence one needs to rotate the gravity gradient tensor. This is performed by applying a product of three orthogonal rotations, starting with the rotation along $y$, whose amplitude eventually largely dominates in Nadir pointing. As long as the two other rotations are small enough, this correctly deals with the influence of the satellite rotation at the leading orders.

We also consider to implement compensation systems for the two following physical quantities:

- the rotation of the satellite, in order to obtain mirrors with a fixed orientation in the frame of the atoms. This configuration can be obtained by tilting the two first and last retroreflecting mirrors by angles $\pm\theta=\pm \Omega_y T$, where $\Omega_y$ is the rotation rate along $y$, the cross-track axis, as displayed in figure~\ref{tiltedmirrors}. As we will see below, this configuration removes the sensitivity of the interferometer to centrifugal and Coriolis accelerations;
\begin{figure}[ht]\centering 
\includegraphics[width=5in]{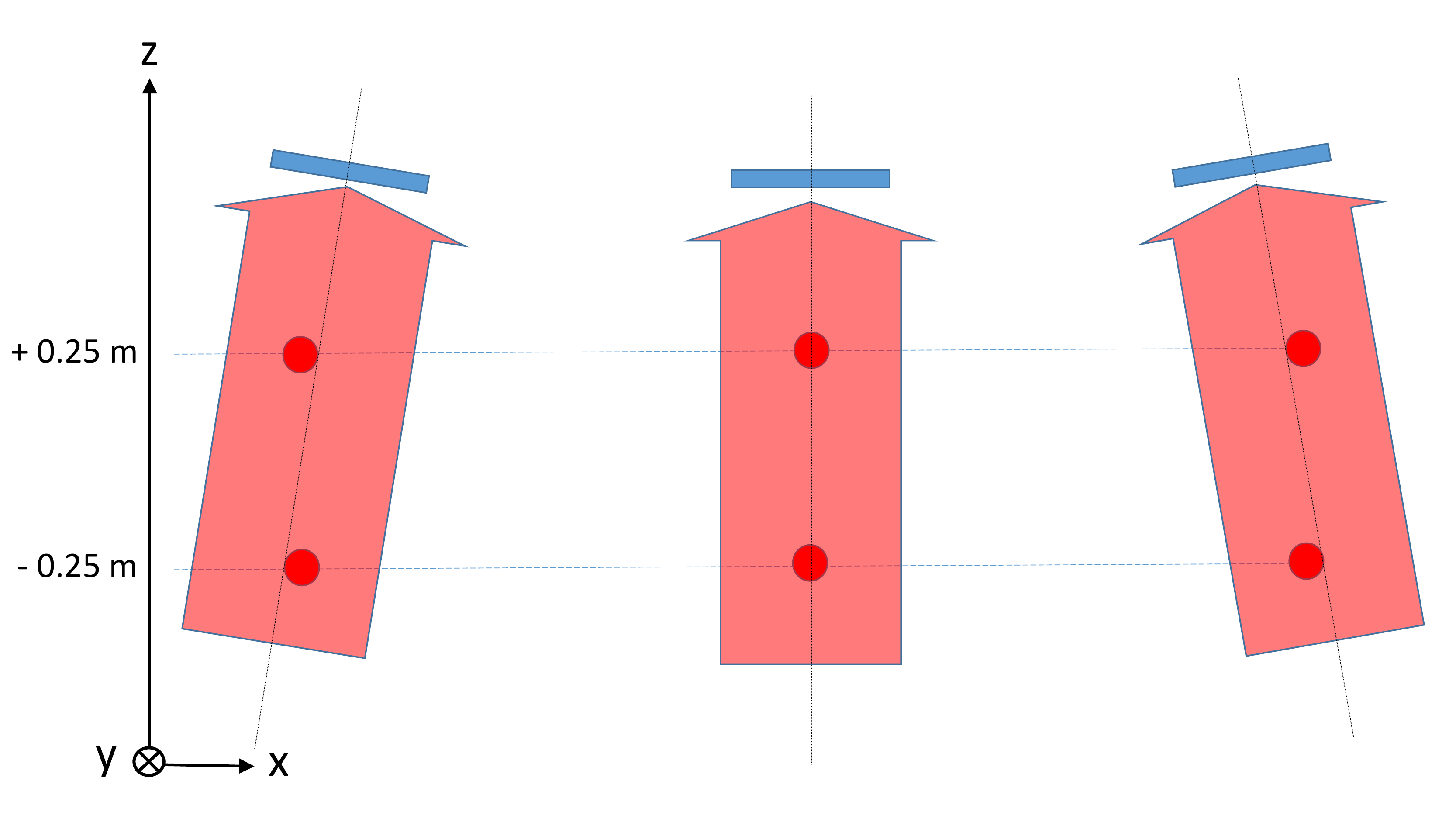}
\caption{Tilted mirror configuration.}
\label{tiltedmirrors}
\end{figure}

- the phase shift induced by the gravity gradient, as recently proposed in \cite{Roura2017}. This relies on an adequate change by $\delta k$ of the Raman wavevector at the second pulse. This method has recently been demonstrated in \cite{Damico2017,Overstreet2018}.

The trajectories of the atomic wavepackets along the arms of the interferometer are determined analytically by solving the Euler-Lagrange equations using a power series expansion as a function of time $t$, as in ~\cite{Hogan2009}. Subsequently, the phase shift can be calculated out of the positions of the center of the atomic wavepackets at the different Raman laser pulses using the following formula~\cite{Borde2004}
\begin{eqnarray}
\phi = 2k_1\cdot r_A-4k_2\frac{r_B+r_C}{2}+2k_3\frac{r_D+r_{D2}}{2}.
\label{formula1}
\end{eqnarray}
We consider here a $\pi/2-\pi-\pi/2$ double diffraction interferometer, such as displayed later on figure \ref{Interferometer_Scheme}, with effective wave numbers $k_{1,2,3}$, corresponding to two photon transitions. (A, B, D) (resp. (A, C, D2)) are the positions of the centers of the partial wavepackets at the time of the three pulses along the upper (resp. lower) arm of the interferometer.  
This formula is valid for Hamiltonians at most quadratic in position and momentum \cite{storey_1994,Borde2004}, which is the case considered here; terms beyond this approximation can be treated as proposed in \cite{dagostino_2011,Bertoldi2019}.

Table \ref{tab:phasesingle1} presents the dominant terms contributing to the phase of a single interferometer and to the separation between the wavepackets at the output of the interferometer, listed with respect to their scaling on initial coordinates and velocities of the atoms in the satellite frame $(x,y,z,v_x,v_y,v_z)$. Only the dominant terms, which scale as $T^2$ or $T^3$ depending on the terms, are listed here. For the sake of simplicity, we take here $\Omega_x=\Omega_z=0$. We consider the case where the rotation rates $\Omega_{\mathrm{orb}}$, $\Omega_y$ and $\Omega_m$, the equivalent rotation rate of the mirrors given by $\Omega_m=\theta/T$, are different, and the cases where the rotation and/or gravity gradient are compensated ($\Omega_m=\Omega_y$ and/or $\delta k=kT_{zz}T^2/2$). 

Here, $T$ denotes the free evolution time between pulses, $k = 4\pi/(780 \mathrm{nm})$ the effective wave number for a two-photon Raman transition, $T_{zz}$ the gravity gradient. Typical values for the relevant parameters are $\Omega_{\mathrm{orb}} = 1.2\,\mathrm{mrad/s}$ (for an altitude of about 250 km), $T = 5\,\mathrm{s}$, $T_{zz} = -2.7\cdot10^{-6}\,\mathrm{s}^{-2}$ and $k=16.1\cdot10^{6}\,\mathrm{m}^{-1}$.

The gravity gradient phase shift is given by $\Phi=2kzT_{zz}T^2$, and thus leads to a differential phase of $\Delta\Phi=2kDT_{zz}T^2$ between two interferometer separated along $z$ by the distance $D$. For $D=0.5$~m, and the parameters above, this phase shift is as large as 1087~rad. We assume that the individual phase measurements are performed with a sensitivity limited by detection noise at the quantum projection limit, for which $\sigma_\Phi=1/\sqrt{N}$, $N$ being the number of detected atoms, assuming an ideal contrast. For $10^6$ detected atoms at the output of each interferometer, we obtain an expected sensitivity of $\sigma_{T_{zz}}=\sqrt{2}/2\sqrt{N}kDT^2=3.5$~mE/shot. 

 \begin{table*}[!ht]
 \begin{center}
 \caption{Leading terms in the phase of a single interferometer}
 \small
 \begin{tabular}{|p{0.12\textwidth}|p{0.25\textwidth}|p{0.25\textwidth}|p{0.25\textwidth}|}\hline
    & \textbf{General case} & \textbf{Compensated rotation} & \textbf{Compensated gradient and rotation, Nadir pointing}  \\ \hline
	 & Any $\Omega_{orb,y,m}$ & $ \Omega_m=\Omega_y$ &$ \Omega_m=\Omega_y=\Omega_{\mathrm{orb}}$ \\ 
	 & $\delta k=0$ & $\delta k=0$ &$\delta k=kT_{zz}T^2/2$ \\ \hline
 $x$ & $kx(T_{xx}(5\Omega_y-4\Omega_m-\Omega_{\mathrm{orb}})+4(\Omega_y-\Omega_m)\Omega_y^2+T_{zz}(\Omega_{\mathrm{orb}}-\Omega_y))T^3 $& $kx(T_{xx}-T_{zz})(\Omega_{y}-\Omega_{\mathrm{orb}})T^3 $ & $ 0 $ \\ \hline
 $y$ & 0 & 0 & 0\\  \hline
 $z$  & $2kz(T_{zz}-\Omega_m^2+\Omega_y^2)T^2$ & $2kzT_{zz}T^2$  & 0 \\ \hline
 $v_x$ & $4kv_x(-\Omega_m+\Omega_y)T^2$ & 0 & 0 \\ \hline 
 $v_z$ & $2kv_z(T_{zz}-\Omega_m^2+4\Omega_m\Omega_y-3\Omega_y^2)T^3$ & $2kv_zT_{zz}T^3$ & 0 \\  
 $v_y$ & 0 & 0 & 0\\  \hline
 Separation   &  &  &   \\ \hline
$\Delta x$ & $4 \hbar k T^2(\Omega_m-\Omega_y)/m_{\mathrm{Rb}} $ & 0 & 0  \\
$\Delta y$ & 0 & 0 & 0  \\

$ \Delta z$ & $2\hbar kT^3(T_{zz}-(\Omega_m-3\Omega_y)(\Omega_m-\Omega_y))/m_{\mathrm{Rb}}$ & $2 \hbar kT^3 T_{zz}/m_{\mathrm{Rb}}$ &  0 \\
\hline
\end{tabular}
 \label{tab:phasesingle1}
 \end{center}
 \end{table*}

We first discuss the case where $\Omega_m=0$, i.e. without rotation compensation. The gravity tensor measurement is biased by a contribution in $\Omega_y^2$ due to centrifugal accelerations. The interferometer phase features a Sagnac phase term $4 \, kv_x \, \Omega_y \, T^2$, and the separation between the two partial wavepackets at the output of the interferometer along $x$ is $4 \, \hbar k \, T^2 \, \Omega_y/m_{\mathrm{Rb}}$. This leads to a reduction of the interferometer contrast due to dephasing when averaging the Sagnac phase across the velocity distribution. Equivalently, the contrast is reduced when the separation is comparable to the coherence length of the atomic wavepackets. For the resulting loss of contrast to be negligible, one needs $\sigma_v\ll 1/ \left ( 4k\Omega_yT^2 \right )$. This corresponds to temperatures $T\ll T_0$, where $T_0$ is given by 
\begin{equation}
T_0=\frac{m_{\mathrm{Rb}}}{16k_Bk^2\Omega_y^2T^4} \,
\end{equation} 
where $k_B$ is the Boltzmann's constant. For Nadir pointing, in which $\Omega_y=\Omega_{\mathrm{orb}}=1.2\,\mathrm{mrad/s}$, a temperature lower than $T_0=3$ fK is required, which is well below what can be achieved with current technology. This limit would also apply for measurements along the $x$-axis, which is also impacted by the large rotation rate along the $y$-axis, but not for measurements along the $y$-axis. Thus, as such, gravity gradient measurements in Nadir configuration can only be performed along one axis. 

Also, in a similar manner, gravity gradients put an additional requirement on the atomic temperature, given by $T < T_1$, where $T_1$ is given by 
\begin{equation}
T_1=\frac{m_{\mathrm{Rb}}}{4k_bk^2T_{zz}^2T^6} \, .
\end{equation}    
For $T_{zz} = -2.7\cdot10^{-6}\,\mathrm{s}^{-2}$, $T_1\sim100$~pK. Such a limit is compatible with the ultralow atomic temperatures reached thanks to Delta Kick collimation techniques (of the order of a few tens of pK). This requirement can also be relaxed if the gravity gradient phase shift is compensated for using an appropriate change of Raman laser frequency at the second pulse.

Measuring the three axes in Nadir configuration requires a compensation scheme for the large rotation along $y$, which can be realized by tilting the Raman mirrors, as discussed above. This corresponds to the second case in table \ref{tab:phasesingle1}, where the Sagnac phase, as well as the centrifugal acceleration, are canceled if the angles set on the mirrors are perfectly tuned ($\Omega_m=\Omega_y$).
On the other hand, if the pointing of the satellite is inertial (and no rotation compensation is applied), rotation rates up to $\Omega_0=6\,\mu\mathrm{rad/s}$ can be accepted, for the same temperature limit of 100 pK. This makes gravity measurements along three orthogonal directions possible in inertial pointing mode (with flat mirrors). Equivalently, this translates into the same limit for the maximum rotation rate mismatch between $\Omega_m$ and $\Omega_y$ in the case of imperfect rotation compensation in the Nadir pointing mode: $\delta\Omega=\Omega_m-\Omega_y \ll \Omega_0$. 

Finally, the last column in table \ref{tab:phasesingle1} shows that for a properly tuned rotation rate for Nadir operation ($\Omega_{y}=\Omega_{\mathrm{orb}}$), for properly tilted mirrors ($\Omega_m=\Omega_y$), and for a properly set change in the Raman wavevector at the second Raman pulse ($\delta k=kT_{zz}T^2/2$), all higher order terms in the interferometer phase, as well as the separation of the two wavepackets at the output of the interferometer, are canceled. The loss of contrast due to dephasing is thus suppressed.

\section{Error budgeting}
\label{sec:error}
Having discussed the constraints set by the finite coherence length of the atomic wavepacket onto the contrast of the interferometer, we now examine the requirements on the atomic source parameters, on the Raman laser setup and on the control of rotations to keep the uncertainties in the determination of systematics in the differential measurement below 1 mrad (which corresponds to an error of 3.5 mE).

\subsection{Requirements}

We start by briefly discussing the inertial case. There, the two measurement axes $x$ and $z$, fixed in the satellite frame and lying in the orbital plane, rotate with respect to the frame where the gravity tensor is diagonal. This leads to a mixing between $T_{zz}$ and $T_{xx}$ components, and a modulation of the gradiometer phase, which is given by: $\Phi=2kL(T_{zz}\textrm{cos}(\chi)+T_{xx}\textrm{sin}(\chi))T^2$, where $L$ is the separation between the two interferometers and $\chi$ is the satellite angle position in the orbital plane. $T_{zz}$ and $T_{xx}$ can then be separated by combining the measurements along two orthogonal directions. There, the uncertainty in the determination of the pointing direction $\delta\theta$ leads to an error in the determination of the tensor component $T$ of interest of the order of $\delta\theta \times T$, which amounts up to 3 mE for $\delta\theta=1~\mu$rad. This is a very tight requirement for the mission, especially for a low altitude orbit. The potential of such a configuration for gravity field recovery has already been studied in \cite{Douch2018}, together with the single cross-track axis in Nadir configuration. We thus focus through the rest of the paper onto the case of a 3-axis determination in Nadir configuration, with the help of compensated rotation.   

\begin{table*}[!ht]
 \begin{center}
 \caption{Dominant terms in the residue of the phase of a single interferometer, for compensated gravity gradient and non perfect rotation compensation, corresponding sensitivity for differential measurements, and phase dispersion}
 \small
 \begin{tabular}{|p{0.12\textwidth}|p{0.32\textwidth}|p{0.28\textwidth}|p{0.16\textwidth}|}\hline
    & \textbf{Terms} & \textbf{Differential}  & \textbf{Phase} \\ 
  &  & \textbf{Phase (in rad)} & \textbf{dispersion}\\ 
  &  &  & \textbf{(in rad)}\\ 
  
	  & $\Omega_m=\Omega_y+\delta\Omega_m$  & $\delta\Omega_m=10^{-6}$~rad/s&\\ 
	  & $\delta k=kT_{zz}T^2/2$ & $\Omega_y-\Omega_{\mathrm{orb}}=10^{-6}$~rad/s&\\  
	  &  & $\delta x=10^{-6}$~m & $\sigma_x=0.1$~mm \\ 
	 &  & $\Delta z=L=0.5$~m & $\sigma_z=0.1$~mm\\ 
	  &  & $\delta v_x=10^{-6}$~m/s & $\sigma_v=98~\mu$m/s\\ 
	  &  & $\delta v_z=10^{-6}$~m/s & $\sigma_v=98~\mu$m/s\\ \hline
	
	 $x$ & $-kx((T_{xx}-T_{zz})(\Omega_{\mathrm{orb}}-\Omega_{y})+4\delta\Omega_m (T_{xx}+\Omega_y^2))T^3 $& $-8.3\times10^{-9}$& $-8.3\times10^{-7}$\\ \hline
 $z$  & $-4kz\delta\Omega_m\Omega_yT^2$ & -0.944 & $-1.9\times10^{-4}$\\ \hline
 $v_x$ & $-4kv_x\delta\Omega_mT^2$ &  $-1.6\times10^{-3}$ & -0.157 \\ \hline 
 $v_z$ & $4kv_z\delta\Omega_m\Omega_yT^3$ & $9.4\times10^{-6}$ & $9.2\times10^{-4}$ \\  \hline 
 Separation   &   & \textbf{Separation}&\\ 
    &   & \textbf{(in m)}&\\ \hline
$ \Delta x$ & $4 \hbar k \delta\Omega_m T^2/m_{\mathrm{Rb}} $ & $1.2\times10^{-6}$&\\
$\Delta z$ & $4\hbar k\Omega_y\delta\Omega_m T^3/m_{\mathrm{Rb}} $  & $6.8\times10^{-9}$&\\
\hline
\end{tabular}
 \label{tab:phasesingle3}
 \end{center}
 \end{table*}

Table \ref{tab:phasesingle3} lists the dominant terms in the development of the output interferometer phase in the case where the compensation of the rotation is not perfect ($\Omega_m \neq \Omega_y$). We find for a mismatch of $\delta\Omega_m=10^{-6}$ rad/s, (which corresponds to an error in the tilt of the mirrors of $\delta \theta=\delta \Omega \, T=5\mu$rad) a phase error on the differential acceleration of -944 mrad, due to residual centrifugal accelerations. To keep the error below 1 mrad, an uncertainty in the knowledge of the rotation rate along y of 1 nrad/s is thus required, or equivalently an uncertainty in the tilt (of the two mirrors) of 5 nrad. 

\begin{table}[!ht]
 \begin{center}
 \caption{Requirements on the relative initial positions and velocities for the phase error to remain below 1 mrad, in the case of non perfect rotation compensation $\delta\Omega_m=10^{-6}$~rad/s and non perfect Nadir pointing $\Omega_y-\Omega_{\mathrm{orb}}=10^{-6}$~rad/s}
 	\begin{tabular}{|c|c|}\hline
$x$	  &  $\delta x < 0.12$~m \\ 
$z$	 &   $\delta z < 0.5$~mm\\ 
$v_x$	  &   $\delta v_x < 0.6~\mu$m/s\\ 
$v_z$	  &   $\delta v_z < 106~\mu$m/s\\ 
\hline
\end{tabular}
 \label{tab:phasesingle3b}
 \end{center}
 \end{table}

Table \ref{tab:phasesingle3b} gives the requirements on the source parameters (velocities and positions) to keep the phase error below 1 mrad, in the case of non perfect compensation of the rotation, and for non perfect Nadir pointing. We assume here a knowledge of the rotation rate along y at the level given above. The corresponding requirements are found to be largely manageable. One should nevertheless keep in mind that these requirements scale with the amplitudes of the mismatches $\delta\Omega_m$ and $\Omega_y-\Omega_{\mathrm{orb}}$.

\subsection{Measurements along the two other axes}

The interferometer configuration required to measure $T_{xx}$ is identical to the one studied before, as for any measurement axis in the orbital plane, the rotation rate along $y$ needs to be compensated. The conclusions and requirements derived in the previous sections thus apply, provided the role of $x$ and $z$ are exchanged. As for the cross-track axis, fixed and parallel mirrors can be used, which simplifies the laser setup design and relaxes the constraints on the control of the beam alignment. This was the configuration considered in \cite{Douch2018}.

\subsection{Combining the three signals}

While the requirements on the control and knowledge of $\Omega_{x,z}$ can be met with current technologies, using for instance fiber optic gyroscopes of the Astrix class, the ones of the rotation rate $\Omega_y$, and of the mismatch of the mirror with respect to the ideal tilt, are very stringent, and cannot presently be met, even with the best space qualified gyroscopes. Instead we propose to use the mathematical properties of the gravity tensor, and its null trace, to estimate $\Omega_y$, or at least its fluctuations.

The phase signals of the three interferometers are given by (we consider here only the leading terms, the gravity tensor terms, the gravity gradient compensating terms due to the wavevector change at the second pulse and the centrifugal terms): 
\begin{eqnarray*}
\delta\Phi_x & = & 2kz(T_{xx}-T_{xx}^{\mathrm{eff}}-\Omega_{m,x}^2+\Omega_y^2+\Omega_z^2)T^2\\
\delta\Phi_y & = & 2kz(T_{yy}-T_{yy}^{\mathrm{eff}}+\Omega_x^2+\Omega_z^2)T^2\\
\delta\Phi_z & = & 2kz(T_{zz}-T_{zz}^{\mathrm{eff}}-\Omega_{m,z}^2+\Omega_y^2+\Omega_x^2)T^2
\end{eqnarray*}
with $T_{ii}^{\mathrm{eff}}=2\delta k_{ii}/kT^2$ and $\delta k_{ii}$ the change in $k$ applied in the direction $i$.

Assuming the $T_{ii}^{\mathrm{eff}}$ are tuned so as to null the output phases, we have
\begin{eqnarray*}
T_{xx}^{\mathrm{eff}} & = & T_{xx}-\Omega_{m,x}^2+\Omega_y^2+\Omega_z^2\\
T_{yy}^{\mathrm{eff}} & = & T_{yy}+\Omega_x^2+\Omega_z^2\\
T_{zz}^{\mathrm{eff}} & = & T_{zz}-\Omega_{m,z}^2+\Omega_y^2+\Omega_x^2
\end{eqnarray*}

Summing the three equation and exploiting the null trace relation $T_{xx}+T_{yy}+T_{zz}=0$, we find
\begin{equation}
\sum T_{ii}^{\mathrm{eff}} = -\Omega_{m,x}^2-\Omega_{m,z}^2+2(\Omega_x^2+\Omega_y^2+\Omega_z^2)
\end{equation}

For $\Omega_x$ and $\Omega_z$ well below $10^{-6}$ rad/s (or sufficiently well determined with gyroscopes), and assuming that the mirror tilts are left unchanged, fluctuations of $\Omega_y$ can be determined with an uncertainty limited by the combined sensitivities of the three gradiometers.

Writing $\Omega_y=\Omega_{y0}+\delta\Omega_y$, where $\Omega_{y0}$ is a reference value close to $\Omega_{\mathrm{orb}}$, we obtain
\begin{equation}
\sum T_{ii}^{\mathrm{eff}} \simeq -\Omega_{m,x}^2-\Omega_{m,z}^2+2\Omega_{y0}^2+4\Omega_{y0}\delta\Omega_y
\end{equation}

The uncertainty in the evaluation of $\delta\Omega_y$ is finally given by 
\begin{equation}
\Delta(\delta\Omega_y)=\frac{\sqrt{3} \, \sigma_T}{4 \, \Omega_{\mathrm{orb}}}
\end{equation}
where $\sigma_T$ is the sensitivity of each gradiometer.

For $\sigma_T=3.5$~mE and $\Omega_{\mathrm{orb}}=1.17$~mrad/s, we find $\Delta(\delta\Omega_y)=0.9$~nrad/s.

This determination can in turn be used to correct the measurement along $x$ and $z$ from centrifugal accelerations. Neglecting as before terms related to $\Omega_x$ and $\Omega_z$, this yields the following equations:
\begin{eqnarray*}
T_{xx} & = & T_{xx}^{\mathrm{eff}}+\Omega_{m,x}^2-\frac{1}{2}\sum T_{ii}^{\mathrm{eff}}+\Omega_{m,x}^2+\Omega_{m,z}^2\\
T_{yy} & = & T_{yy}^{\mathrm{eff}}\\
T_{zz} & = & T_{zz}^{\mathrm{eff}}+\Omega_{m,z}^2-\frac{1}{2}\sum T_{ii}^{\mathrm{eff}}+\Omega_{m,x}^2+\Omega_{m,z}^2
\end{eqnarray*}
and finally
\begin{eqnarray*}
T_{xx} & = & \frac{1}{2}(T_{xx}^{\mathrm{eff}}-T_{yy}^{\mathrm{eff}}-T_{zz}^{\mathrm{eff}}+\Omega_{m,x}^2-\Omega_{m,z}^2)\\
T_{yy} & = & T_{yy}^{\mathrm{eff}}\\
T_{zz} & = & \frac{1}{2}(T_{zz}^{\mathrm{eff}}-T_{xx}^{\mathrm{eff}}-T_{yy}^{\mathrm{eff}}+\Omega_{m,z}^2-\Omega_{m,}^2)
\end{eqnarray*}
Exploiting the null trace to correct for the centrifugal acceleration actually also decreases the uncertainty in the gradiometric measurement along $x$ and $z$ by a factor $2/\sqrt{3}=1.15$.

\section{Monte Carlo model of the interferometer}
\label{sec:model}
This section describes the simulations of a gravity gradiometer based on a pair of double diffraction atom interferometers, focusing here on the effects related to the physics of the interferometer, independently from the inertial forces applied to the atoms. A Monte Carlo model of the interferometers was developed in order to precisely evaluate the impact of the experimental parameters, such as related to the lasers or the atomic sources, onto the differential phase between the two interferometers. The interferometers are fed out of a single ultra-cold atomic source which is split using a combination of a Raman and Bragg laser beams into two clouds separated by 50 cm, as represented in figure~\ref{Concept-Scheme}. The two clouds are thus taken to be identical in their initial velocity, temperature, spatial distribution. 

\subsection{Double-diffraction interferometers}
\label{doublediffraction}
The interferometer geometry is based on the double diffraction technique demonstrated in \cite{Leveque2009}. The Raman beams, of wavevectors $\vec{k}_{1}$ and $\vec{k}_{2}$, are brought together onto the atoms before being retroreflected on mirror(s), leading to the existence of two pairs of counterpropagating Raman beams, with opposite effective wavevectors $\pm\hbar\vec{k}_{\mathrm{eff}}$, where $\vec{k}_{\mathrm{eff}}=\vec{k}_{1}-\vec{k}_{2}$. Both pairs are resonant when the motion of the atoms is perpendicular to the laser beams, as no Doppler shift lifts the degeneracy between the resonance condition between them.

Similarly to diffraction by stationary optical waves, the coupling with the two Raman lasers pairs lead to populating several orders of diffraction. In our model, we consider the first 5 coupled atomic states $\vert j\rangle$ ($j=-2\cdots2$). These states correspond to diffraction orders 0, $\pm 1$ and $\pm 2$: they are linked to the interaction of the atoms with 2 or 4 photons following the two directions $\pm\hbar\vec{k}_{\mathrm{eff}}$. They also differ by their internal states (0 and $\pm 2$ correspond to $\vert F=1\rangle$, $\pm\,1$ to $\vert F=2\rangle$) as the Raman pairs couple different electronic states. Then, the evolution of the atomic quantum state during the pulse is calculated by solving the Schr\"odinger equation, generalizing the method of \cite{Moler1992}, based on adiabatic elimination of the excited state.

\begin{figure}[hbtp]
    \centering
    \includegraphics[scale=0.5]{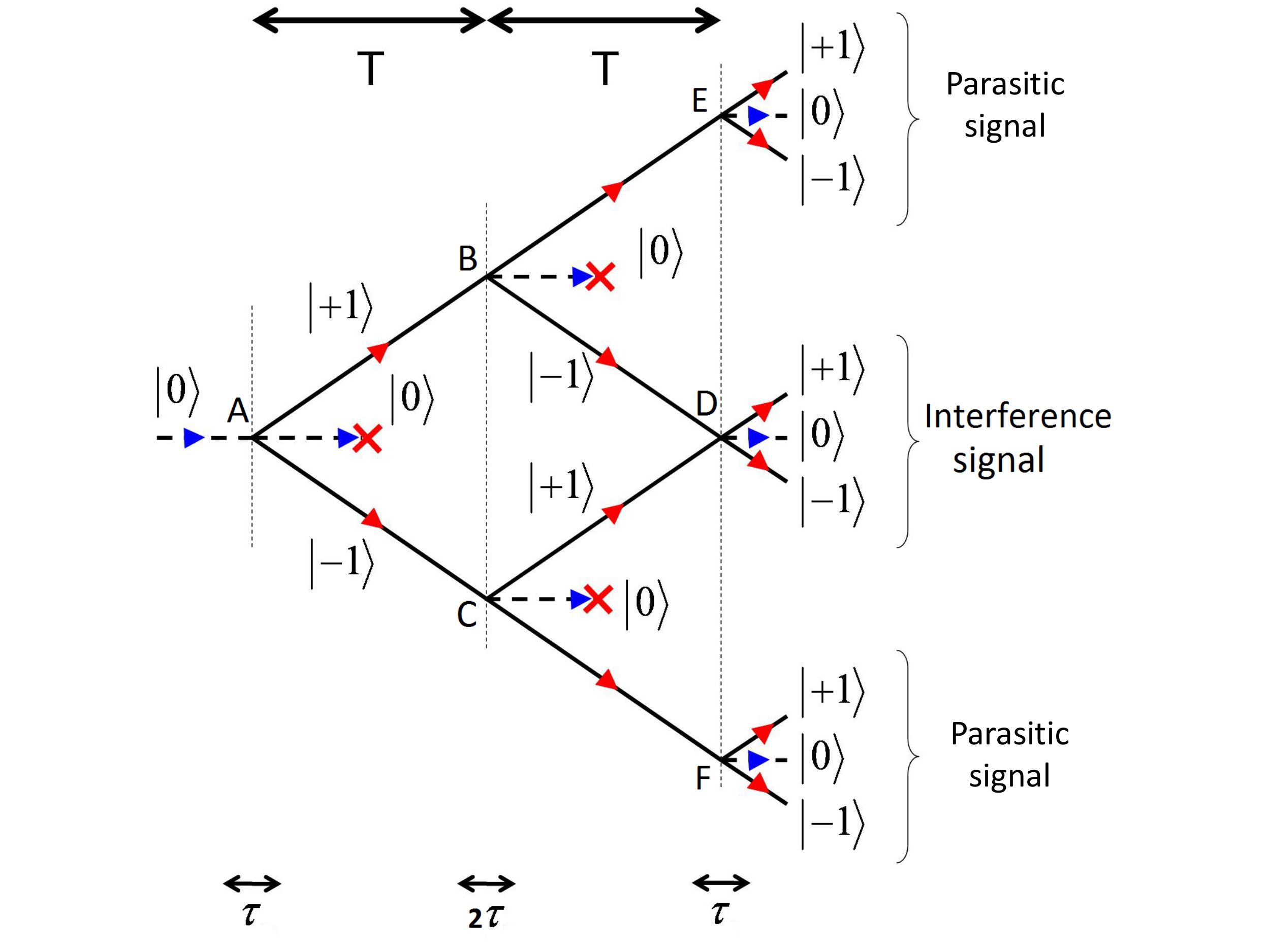}
    \caption{Double diffraction interferometer scheme using three Raman pulses. Note that we do not display the $\vert\pm 2\rangle$ states as they are pushed away together with the $\vert 0\rangle$ wave-packets.}
    \label{Interferometer_Scheme}
\end{figure}
    
The double diffraction interferometer is realized using a three Raman pulses sequence $\frac{\pi}{2} - \pi - \frac{\pi}{2}$, as shown in figure~\ref{Interferometer_Scheme}, separated from each other by a free evolution time $T$. The three pulses are realized with the same laser power but with different duration corresponding respectively to $\tau - 2\tau - \tau$. The duration $\tau$ is defined with respect to the effective Rabi pulsation $\Omega_{\mathrm{eff}}$, using the relation: $\tau =\pi / \left ( \sqrt{2} \, \Omega_{\mathrm{eff}} \right )$, so that the first Raman pulse enables to split the wave-function of the atoms in coherent superposition between the two coupled states $\vert 0\rangle\longrightarrow\vert\pm 1\rangle$. During the first free evolution time $T$, the two arms of the interferometer separate. The second Raman pulse acts on each interferometer arms in order to deflect them (in our case $\vert\pm 1\rangle$ into $\vert\mp1\rangle$). The first and second Raman pulses can also populate other coupled states (such as $\vert\pm 2\rangle$), leading to parasitic paths which could interfere. We suppress these unwanted paths which are in the $\vert F=2\rangle$ state by pushing them away of the interferometer area using resonant laser beams, after each Raman pulse, as shown in figure~\ref{Interferometer_Scheme}. The non-deflected wave-packets after the middle pulse do not disturb the measurement process if we detect only the ``interference signal'' (see figure~\ref{Interferometer_Scheme}), which is possible in our configuration due to the large distance between the different wave-packet trajectories. 

After the second $\pi$-Raman pulse, the two wave-packets $\vert\mp1\rangle$ get closer and overlap after a second free evolution time $T$. Finally, the last $\frac{\pi}{2}$ Raman pulse recombines these atomic wave packets, realizing thus a Mach-Zehnder type interferometer. 

The interferometer phase, which corresponds to the difference between the phase shifts accumulated by the two interferometer arms, is finally extracted from the measurement of the transition probability $P=N_1/(N_1+N_2)$, where $N_1$ and $N_2$ are respectively the number of atoms detected in the hyperfine states $\vert F=2\rangle$ (corresponding to the atoms in state $\vert 0\rangle$) and $\vert F=1\rangle$ (corresponding to the sum of the atoms in states $\vert \pm 1\rangle$).

With this geometry, the interferometer phase is given by:
\begin{eqnarray}
\Delta\Phi & = & \left[\phi_{\downarrow}(\vec{r}_{A}) - \phi_{\downarrow}(\vec{r}_{C}) + \phi_{\uparrow}(\vec{r}_{C}) - \phi_{\uparrow}(\vec{r}_{D})\right] \nonumber \\
& &  - \left[\phi_{\uparrow}(\vec{r}_{A}) - \phi_{\uparrow}(\vec{r}_{B}) + \phi_{\downarrow}(\vec{r}_{B}) - \phi_{\downarrow}(\vec{r}_{D})\right]
\label{Analytic_PaseShift}
\end{eqnarray}
where $\vec{r}_{A},\vec{r}_{B},\vec{r}_{C}$ and $\vec{r}_{D}$ are the center of mass positions of the atomic wave-packets at the different locations $A, B, C$ and $D$ represented in the figure~\ref{Interferometer_Scheme}, and $\phi_{\uparrow}$ (resp. $\phi_{\downarrow}$) the phase difference between the $k_{\uparrow}$ (resp. $k_{\downarrow}$) pair of Raman lasers. Equation \ref{Analytic_PaseShift} generalizes equation \ref{formula1} and allows to account for differences between the wavevectors of the counterpropagating Raman lasers pairs.

\subsection{Description of the Monte-Carlo simulation}

Using this model for the interferometer, we have developed Monte-Carlo simulations of the space-borne gravity gradiometer depicted in figure~\ref{Concept-Scheme}. We average the contribution to the output signals of a large ensemble of atoms, randomly drawing their initial positions and velocities in Gaussian distributions, and calculating the evolution of their wave-function as well as their classical trajectory along the two interferometer paths. The initial mean longitudinal velocity of the atoms is $4v_{rec}$, and the rms initial atomic position is taken to be 100~$\mu$m. The initial mean vertical velocity, ideally null, can be taken different from zero when we estimate the effect of an initial velocity drift along the direction of the lasers onto the interferometer. By simulating two interferometers at different initial positions, and computing the difference between the output phase shift, we simulate a cold atom interferometer gravity gradiometer. 

To simulate the propagation of the atoms through the interferometer, we consider that the momentum kicks occur at the middle of the Raman pulses and we neglect the variations of their position during the pulses. The duration of the first Raman pulse is $\tau =\frac{\pi}{\sqrt{2}\Omega_{\mathrm{eff}}}$, where $\Omega_{\mathrm{eff}}$ is the Rabi angular frequency at the center of the Raman lasers, whose intensity profiles are Gaussian with identical waists $w_0$. Therefore, the effective Rabi angular frequency $\Omega_{\mathrm{eff}}(\vec{r})$ seen by the atoms, as well as the phase shifts $\phi_{\uparrow\downarrow}(\vec{r})$, depend on their position in the Raman beams $\vec{r}$ at the time of the pulses. 

\subsection{Results of the simulation}

The following section discusses the results from the simulation and derives the requirements.

\subsubsection{Parallel retroreflecting mirrors}

\begin{figure}[hbtp]
\centering
\includegraphics[scale=0.8]{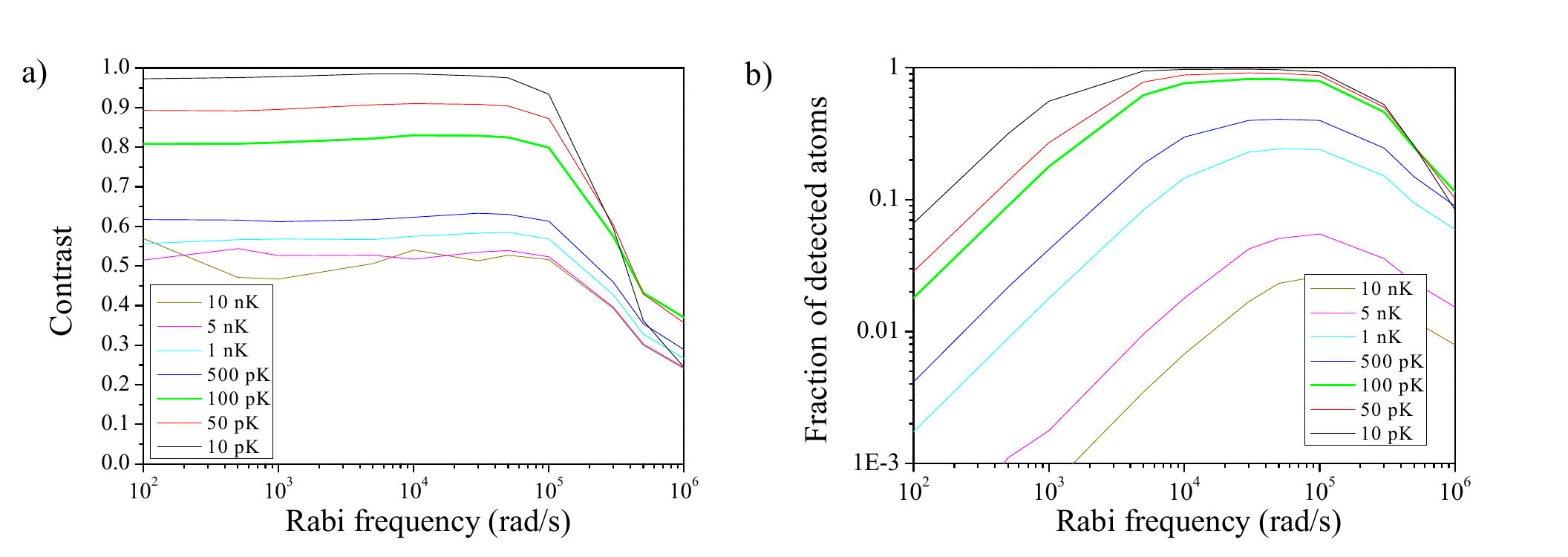}
\caption{Contrast (a) and number of detected atoms (b) as a function of the Rabi pulsation $\Omega_{\mathrm{eff}}$ for different atom temperatures.}
\label{Contrast-NatmDet_Temperature-Omega}
\end{figure}

We start by calculating the contrast and fraction of detected atoms as a function of the Rabi frequency and the temperature, for parallel Raman mirrors (which corresponds to the case where the measurement axis is cross-track, i.e. along $y$). The results are displayed in figure~\ref{Contrast-NatmDet_Temperature-Omega} for temperatures ranging from 0.1~pK to 10~nK, and Rabi pulsation from 5~rad/s to 10~Mrad/s. Here, the Raman laser beams are taken as Gaussian beams with a 5~mm waist size. 

In figure~\ref{Contrast-NatmDet_Temperature-Omega}(a) we present the interferometer contrast 
with respect to $\Omega_{\mathrm{eff}}$ for different temperatures. A plateau of contrast $>0.8$ is observed for a range of Rabi pulsation between $10^{2}$ and $10^{5}$~rad/s and a temperature range between 10 and 100~pK. The corresponding fractions of detected atoms are displayed in figure~\ref{Contrast-NatmDet_Temperature-Omega}(b), where a similar plateau is found, though smaller, corresponding to a range of $\Omega_{\mathrm{eff}}$ between $5.10^{3}$ and $2.10^{5}$~rad/s and a temperature range between 10 and 100~pK. These results confirm the expectations that at large effective Rabi pulsations ($>10^{5}$~rad/s), the coupling to higher momentum states ($\vert\pm 2\rangle$ ...) leads to a loss of contrast. Also, when the temperature increases, the fraction of detected atoms decreases quickly due to the velocity selectivity of the Raman transitions. This motivates to work with the lowest possible temperatures.

Based on these calculations, we select for the rest of the simulations the following parameters: $\Omega_{\mathrm{eff}}=40\times10^{3}$ ~rad/s and a temperature of 100~pK, for which the contrast C is $\approx$ 80\% and the fraction of detected atoms $\approx$ 80\%. Temperatures below 100~pK range have already been obtained using the delta-kick collimation technique~\cite{Kovachy2015PRL}. As for the Rabi pulsation, it corresponds to Raman pulses duration $\tau$ of the order of 55~$\mu$s similar to what is used in standard ground-based Raman interferometers~\cite{Leveque2009}.  

\begin{figure}[hbtp]
\centering
\includegraphics[scale=0.35]{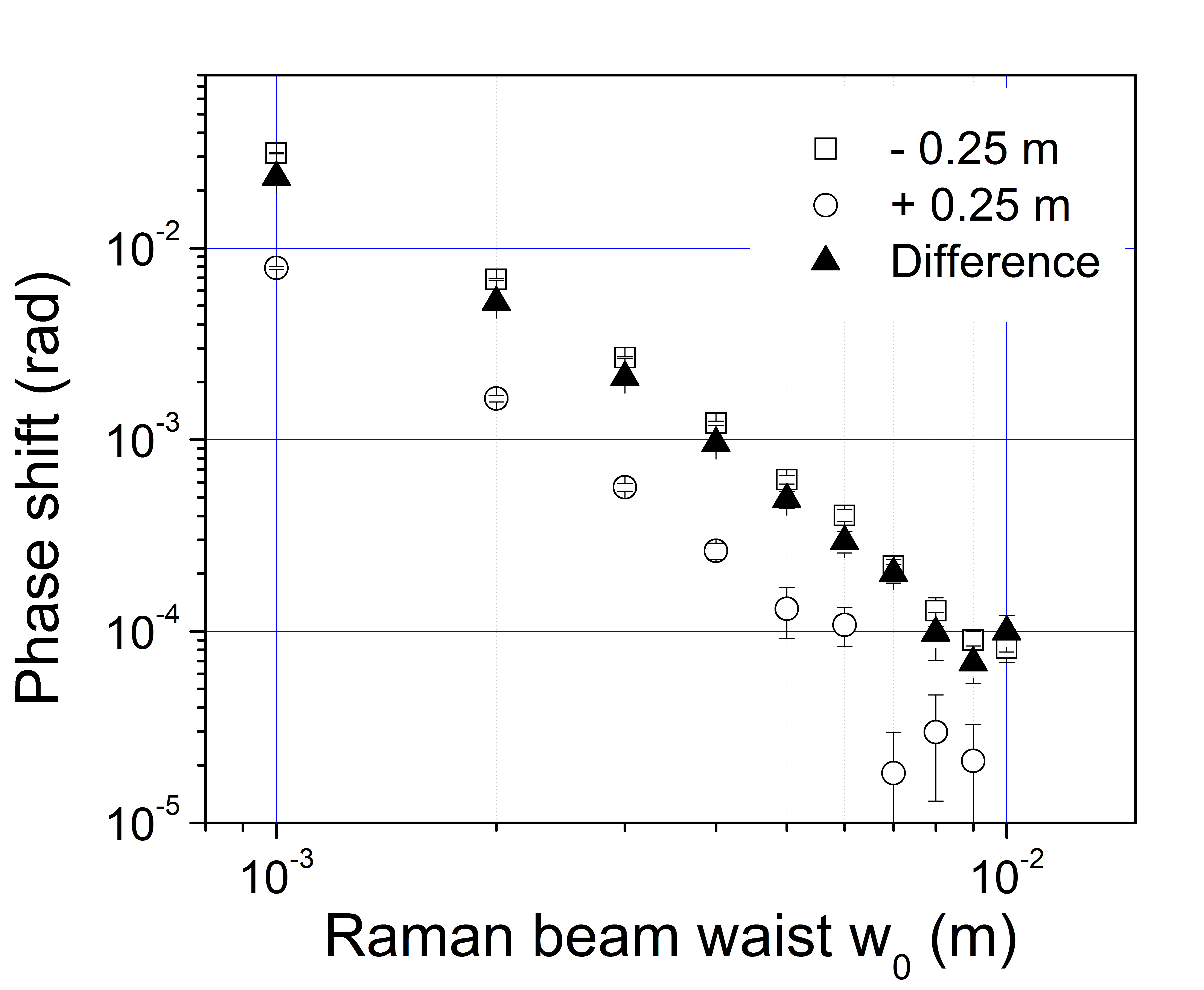}
\caption{Effect of the Raman lasers waist on the phase shifts at the output of the interferometer at 0.25~m $y$-position (open circles) (resp. at -0.25~m $y$-position (open squares)), and on the differential phase shift between the two interferometers (full triangles). All the Raman laser beams have the same size, at the same $y$-position.}
\label{Effect_Laser_Waist_Position}
\end{figure}

The simulation also allows evaluating the effect of the finite size of the Raman beams onto the phase of the interferometers, due to the effects of curvature and Gouy phase. Figure~\ref{Effect_Laser_Waist_Position} shows the calculated phase shifts as a function of the Raman laser waists in the 1-10 mm range, assuming identical waists for the three beams, located at $y=0$, in between the two interferometers at $y=\pm25$~cm. The temperature is $T$ = 100~pK and the effective Rabi pulsation 40~krad/s. The retroreflecting mirrors are placed at $y=+0.4$~m.

Figure~\ref{Effect_Laser_Waist_Position} shows the calculated phase shifts for each individual interferometers, displayed as open squares and circles, and their difference, displayed as full triangles. We find phase shifts that decrease quickly with increasing waist sizes, which are not suppressed in the differential measurement. The resulting systematic effect on the gradiometer phase is lower than 1 mrad for waists larger than 4 mm, and was found to be dominated by the impact of the residual curvature of the wavefront rather than the Gouy phase. This motivates the choice made above of a waist of 5 mm.

The relative positions of the Raman laser waists were then varied in order to evaluate the effect of their positions on the differential phase shift of the gravity gradiometer, and define an error margin for the adjustment of the waist position of the Raman laser beams. The positions of the waists of the three incoming Raman beams were randomly drawn in the range $\pm 10$~m (resp. $\pm 50$~m) with respect to the $y=0$ position at the middle of the two interferometers. The corresponding gradiometer phase shifts were found to vary respectively within $\pm 0.5$~ mrad (resp. within 6 mrad) around the average value of 0.5 mrad. In order to keep the differential phase shift $<$ 1~mrad, the relative $y$-positions of the Raman laser waists should thus be in the range of $\pm$10~m. The Rayleigh length of a Gaussian beam of 5~mm waist being 100 m, this corresponds to a maximum radius of curvature of 1 km at 10~m from the waist, which is well within the measurement capabilities of state of the art wavefront sensors. 

The model was also used to evaluate the impact of other effects, such as light shifts, residual mean Doppler shifts. In particular, we calculated for single interferometers a residual sensitivity to the mean initial velocity along $y$. The sensitivity amounts to 0.05~mrad per $\mu$m/s of mean velocity drift for our parameters (Rabi frequency $\Omega_{\mathrm{eff}}$ = 40~krad/s), and increases when decreasing the Rabi frequency. A control of the relative initial vertical velocity between the two atomic clouds at the input of each interferometer better than 20~$\mu$m/s is thus required to keep the phase error below 1 mrad. 

\subsubsection{Tilted retroreflecting mirrors}

\begin{figure}[hbtp]
\centering
\includegraphics[scale=0.21]{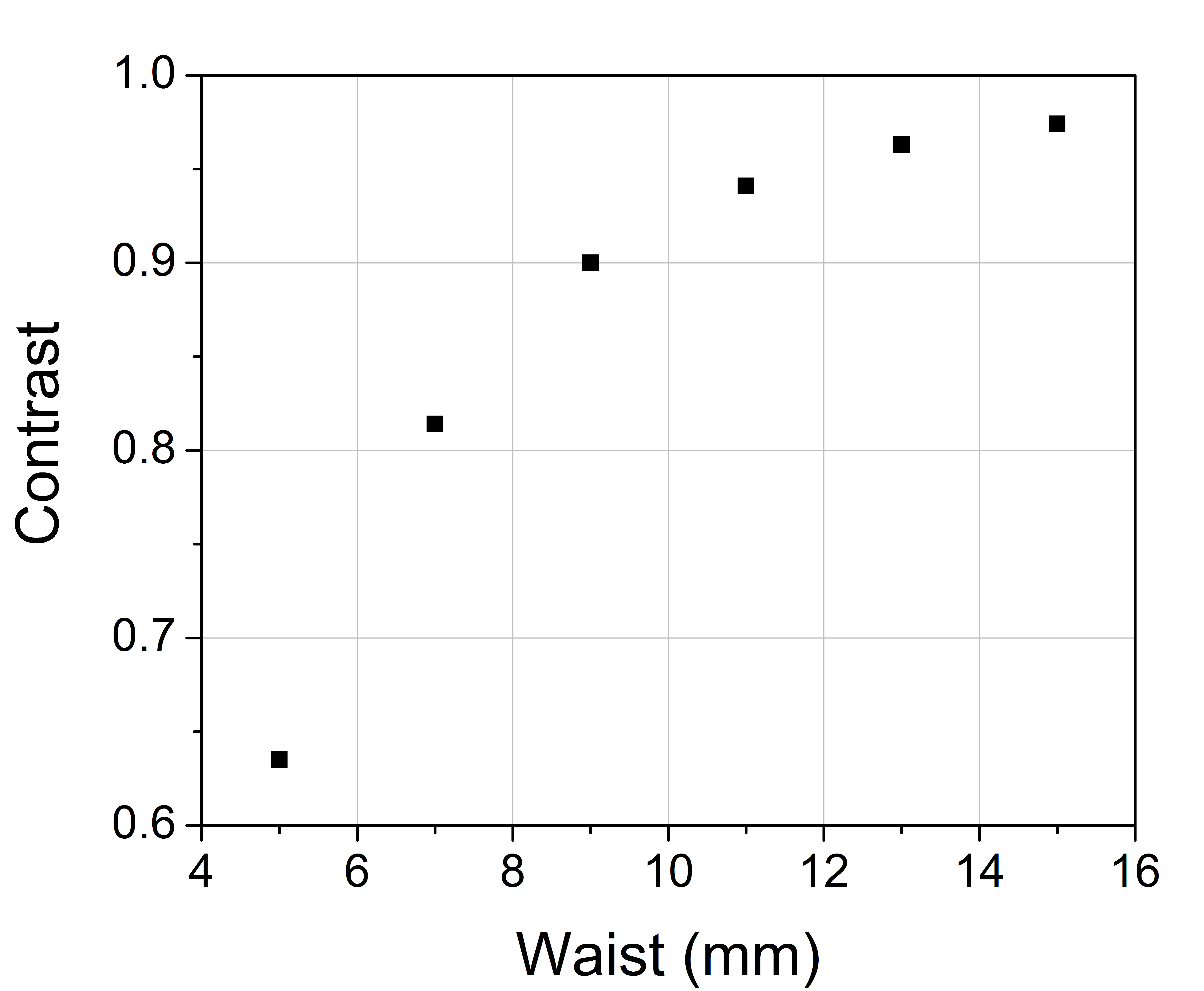}
\includegraphics[scale=0.21]{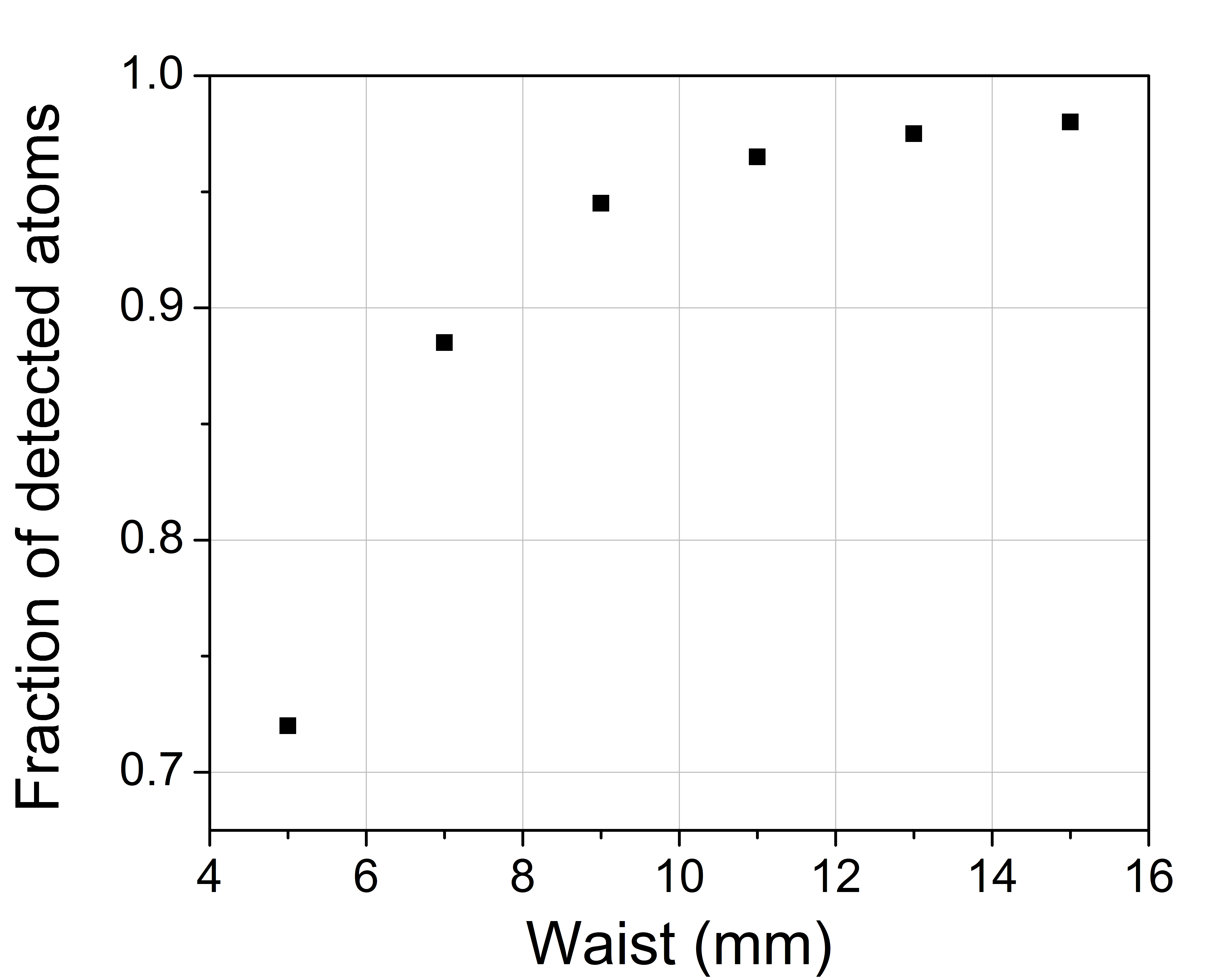}  
\includegraphics[scale=0.21]{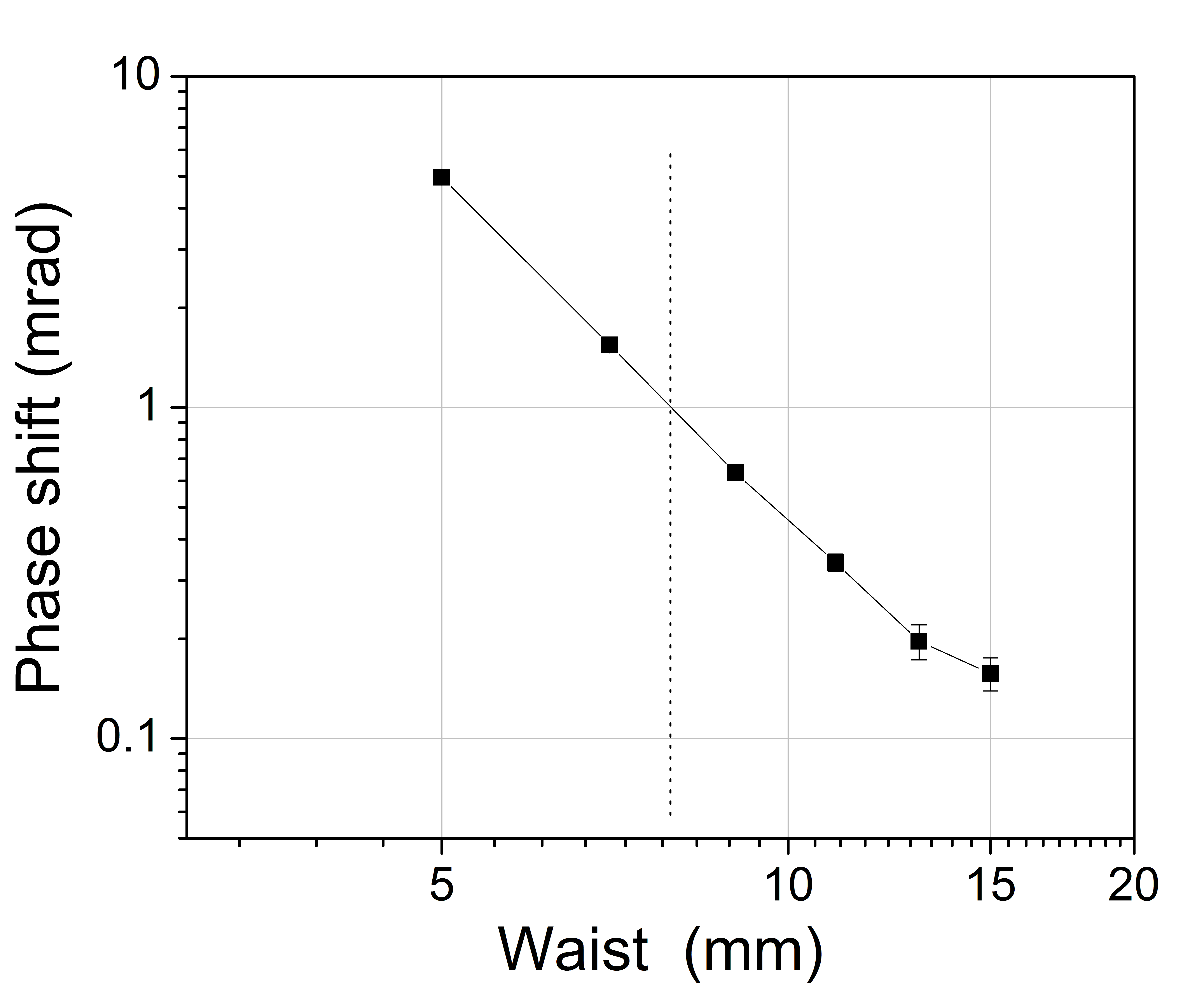}
\caption{Effect of the Raman lasers waist on the contrast (left), fraction of detected atoms (center), and on the differential phase shift between the two interferometers (right), for tilted Raman mirrors. All the Raman laser beams have the same size, at the same $z$-position.}
\label{Effect_Laser_Waist_Position_Tilted}
\end{figure}

In the tilted mirror configuration, finite size effects are expected to have a stronger impact, as the positions of the atomic clouds are (symmetrically) offset with respect to the centre of the Raman beams by about 1.5~mm at the first and third Raman pulses. Figure~\ref{Effect_Laser_Waist_Position_Tilted} presents the calculated contrast (left) and fraction of detected atoms (center) as a function of the beam waist. As expected, smaller contrasts and fraction of detected atoms are found with respect to the parallel configuration (63 \% of contrast instead of 80 \% for a waist of 5 mm). The effect of the curvature onto the gradiometer phase is displayed on figure~\ref{Effect_Laser_Waist_Position_Tilted}-right), where a waist larger than 8 mm is required to keep the phase error below 1 mrad. We finally chose a waist of 1~cm and evaluated the impact of the Rabi frequency and temperature onto the contrast and fraction of detected atoms. We found similar behaviours as before, with plateaus in the trends with respect to the Rabi frequency, with significantly higher contrast (92 \%) and fraction of detected atoms (95 \%) for a temperature of 100 pK. 

\section{Design of the instrument}

\label{Sec:Design}

\subsection{Vacuum system}

\begin{figure}
  \centering
  \includegraphics[width=12cm]{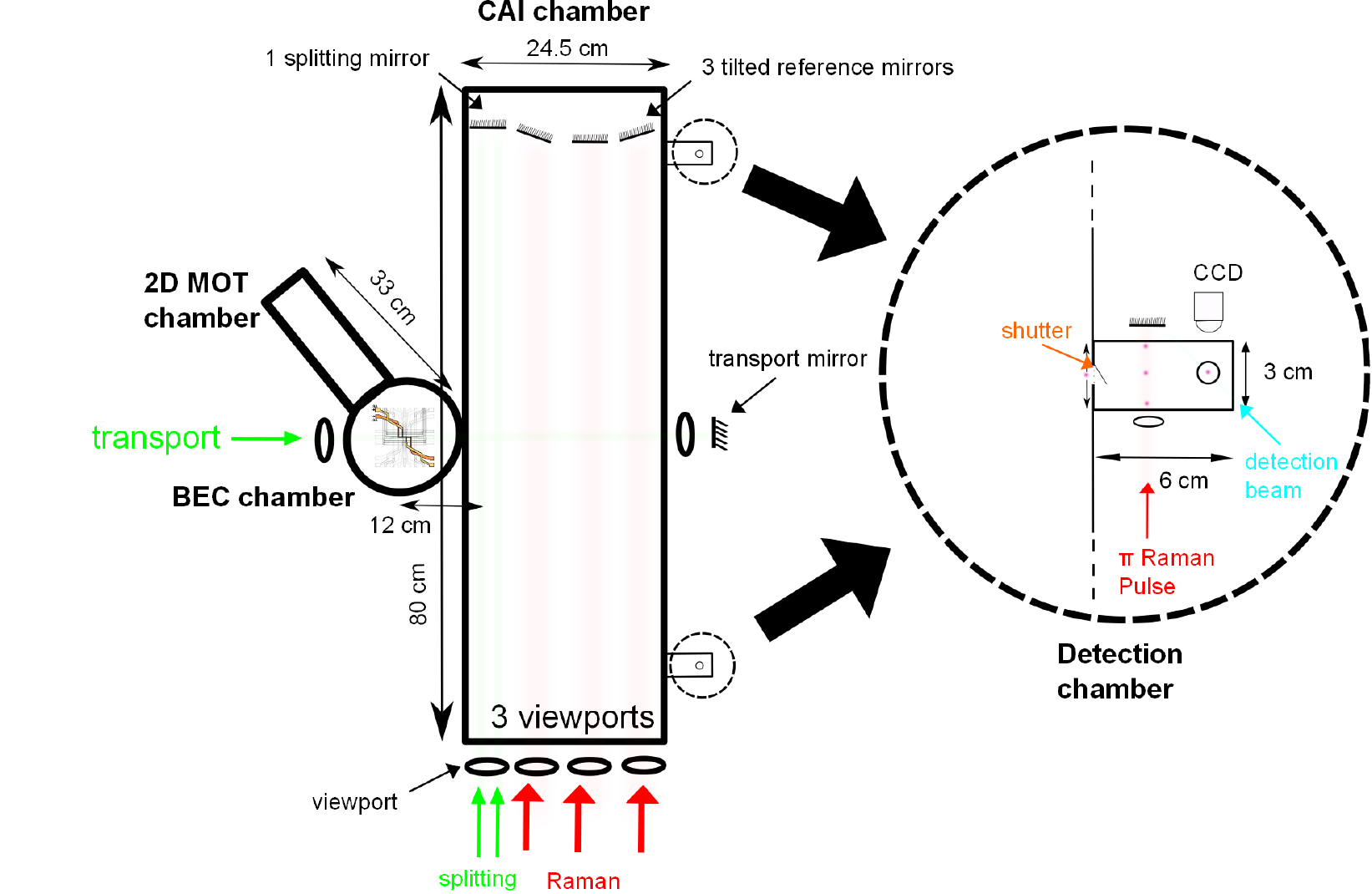}\\
  \caption{Design of the vacuum system for $T_{zz}$ and $T_{xx}$. An atom beam is produced in the 2D MOT chamber and used to load a mirror 3D MOT on a chip in the BEC chamber, where the ultra-cold atom source is achieved. The atom cloud is then launched thanks to a Raman pulse and horizontal Bloch lattices towards the CAI chamber where the differential interferometer is produced. The atom cloud is slowed down thanks to horizontal Bloch lattices, then split and transported at the entrance of the interferometer area by applying vertical Bloch lattices. The detection is achieved in a separated small chamber in order to avoid parasitic light in the CAI interferometric chamber. A $\pi$ Raman pulse is applied 1 s after the last beamsplitter pulse, to have the 3 output ports overlapped 1 s later for counting by fluorescence detection on a CCD camera.}\label{VacuumSchematics}
\end{figure}

Figure~\ref{VacuumSchematics} shows the design of the vacuum system for $T_{zz}$ and $T_{xx}$. The main dimensions of the vacuum chamber are adapted to the displacement of the atomic clouds. The architecture of the BEC chamber, where the ultra-cold atomic source is produced, is inspired by the solution designed for STE-QUEST ATI \cite{Aguilera2014} but without the dipole trap. In the 2D-MOT chamber, a beam of pre-cooled atoms is created by a two dimensional MOT. This pre-cooled beam of atoms is formed out of the background gas pressure created by a reservoir. In the BEC chamber this atom beam is captured by a three dimensional MOT and the atoms are then transferred in a purely magnetic trap. The magnetic fields for the traps are created by a combination of a three-layer atom chip and comparably small magnetic coils. In the magnetic trap, the atomic cloud is compressed and then cooled via RF-evaporation. The chip is parallel to the plane of the atomic clouds displacements. The transport beams launch the atoms into the CAI chamber. 

Four mirrors are fixed inside the vacuum system: one for the two vertical Bloch lattices and three tilted reference mirrors for the interferometer to compensate for the rotation of the satellite which can create bias terms in the output phases and a loss of contrast. The relative angle between two consecutive reference mirrors is $\sim 7$~mrad corresponding to the mean rotation rate of the satellite. The two mirrors for the $\pi/2$ pulses are fixed on piezoelectric tip-tilt mounts to allow the fine control of the relative angle between the three reference mirrors \cite{Hauth2013}. A dynamic range of $\pm 30~\mu$rad and an accuracy of 10 nrad is needed, which is slightly beyond the state-of-the-art of this technology and requires custom development. The impact of these actuators on the power budget is not negligible and their power consumption needs to be optimized. For the $T_{yy}$ CAI, a single reference mirror is used for the splitting and the three interferometer pulses instead of three independent reference mirrors and no tip-tilt mount is needed.

During the interferometer the atom clouds pass through the CAI to finally reach their respective detection chambers. Figure~\ref{VacuumSchematics} zooms in the detection zone. The idea is to wait for the atom cloud to exit the interferometer chamber to avoid parasitic light due to fluorescence. A shutter is placed at the entrance to completely block the scattered photons to reach the CAI chamber. A double diffraction $\pi$ pulse is applied to bring the diffracted states back at the center. Spatial fringes on the atomic population are observed with a CDD camera.

\subsection{Detection signal}

Spatially resolved detection prevents the contrast loss determined by the inhomogeneous dephasing due to initial velocity and position distribution and allows the extraction of information on velocity dependent phase shifts \cite{Dickerson2013PRL,Sugarbaker13PRL}. We consider at first a point-like atomic source, to evaluate the effects of the satellite angular rotation $\Omega_{y,y,z}$ when $\Omega_m=0$ on the final fringe pattern thanks to the ballistic expansion of the atomic ensemble during the long interrogation time. The remaining phase terms are those related to the initial velocities of the atoms ($v_x, v_y, v_z)$, and including Sagnac terms and the effect of the vertical component of the gravity-gradient $T_{zz}$:

\begin{equation}
\phi(v_x, v_y, v_z)= 4 k \left ( v_x \Omega_y +  v_y \Omega_x \right ) T^2 + 2 k v_z \left ( T_{zz} - 3\Omega_y^2 \right ) T^3
\label{eq:phase_velocity}
\end{equation}

\begin{figure}
\centering
\includegraphics[width=0.24\textwidth]{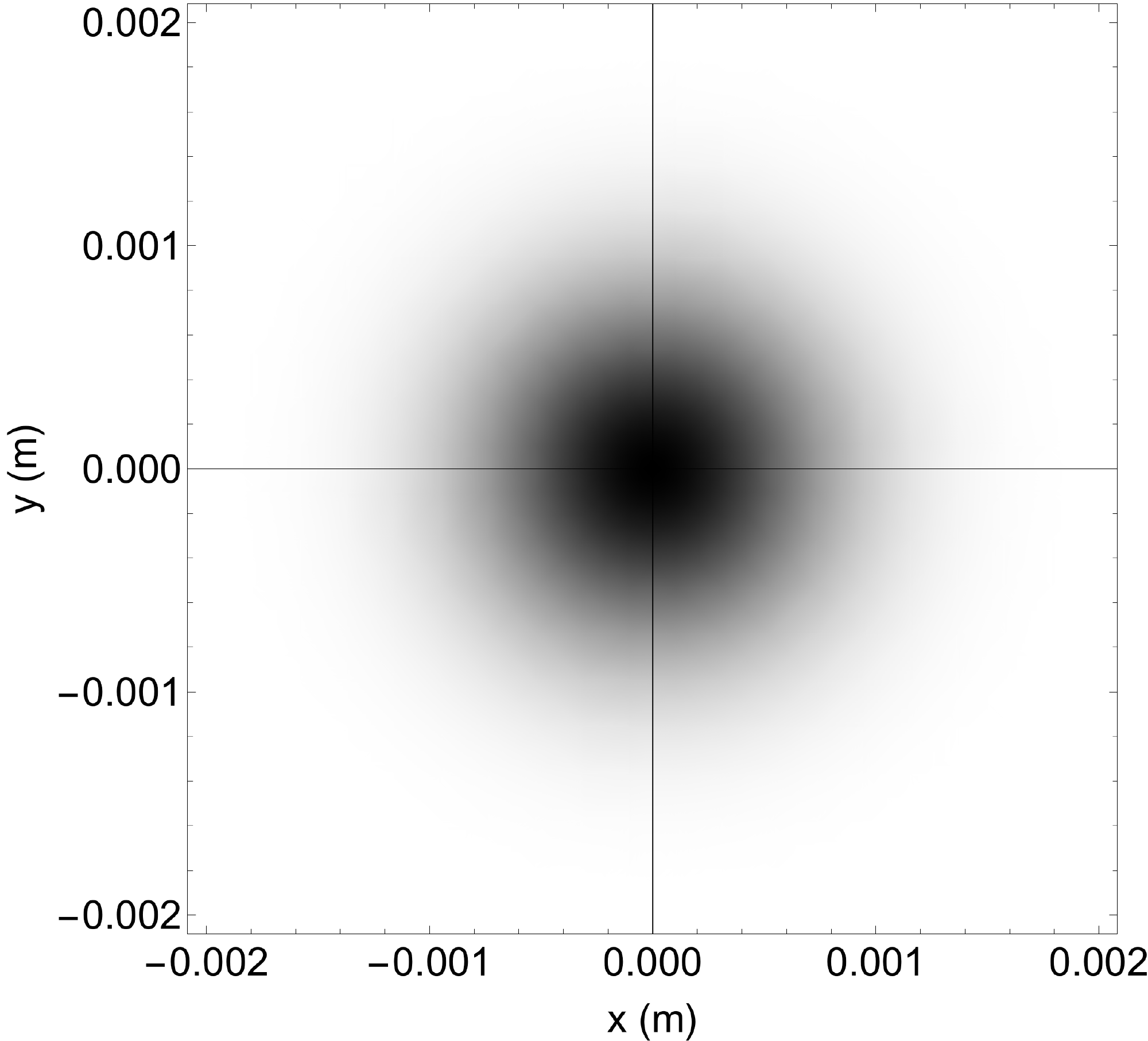}
\includegraphics[width=0.24\textwidth]{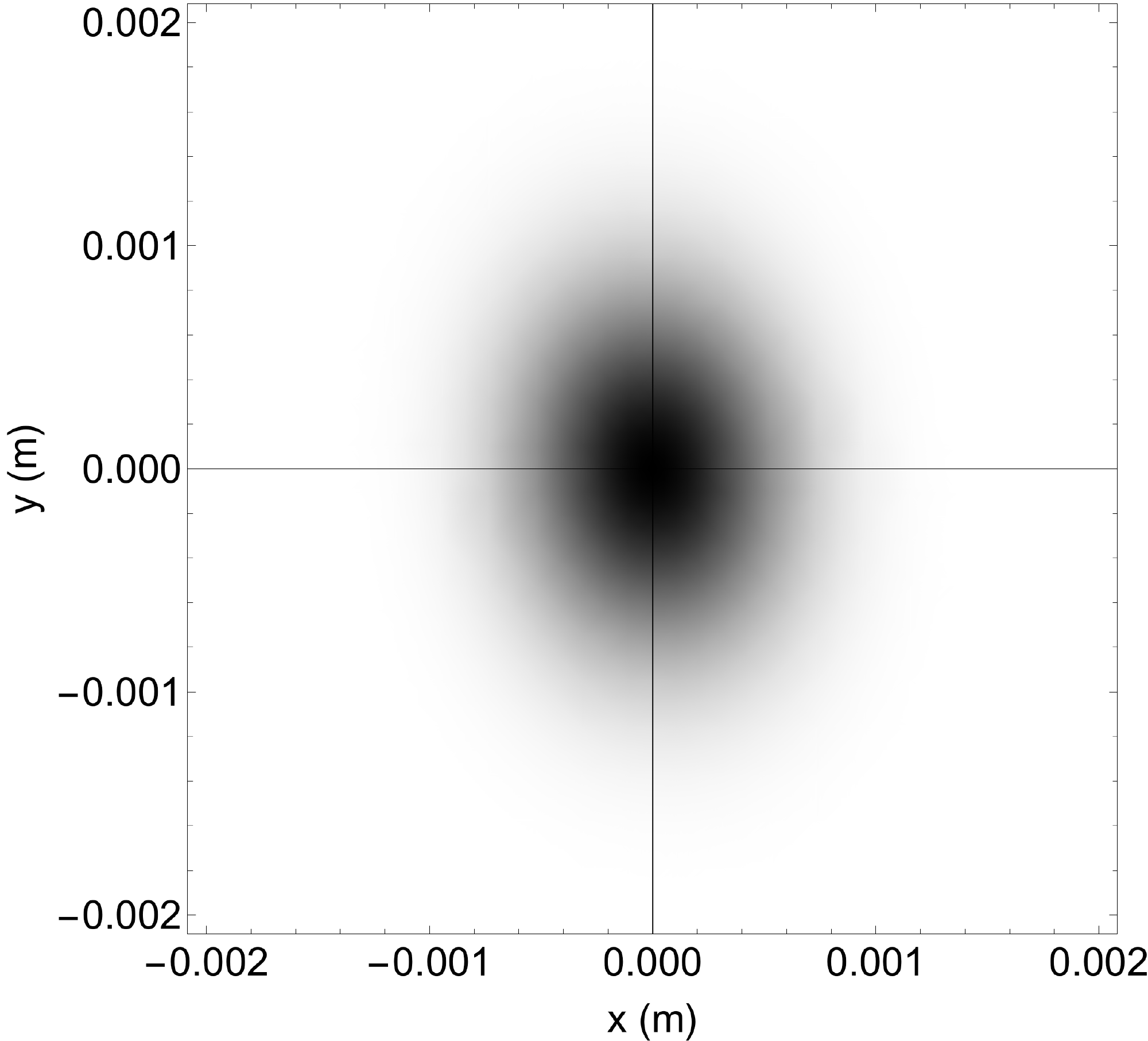}
\includegraphics[width=0.24\textwidth]{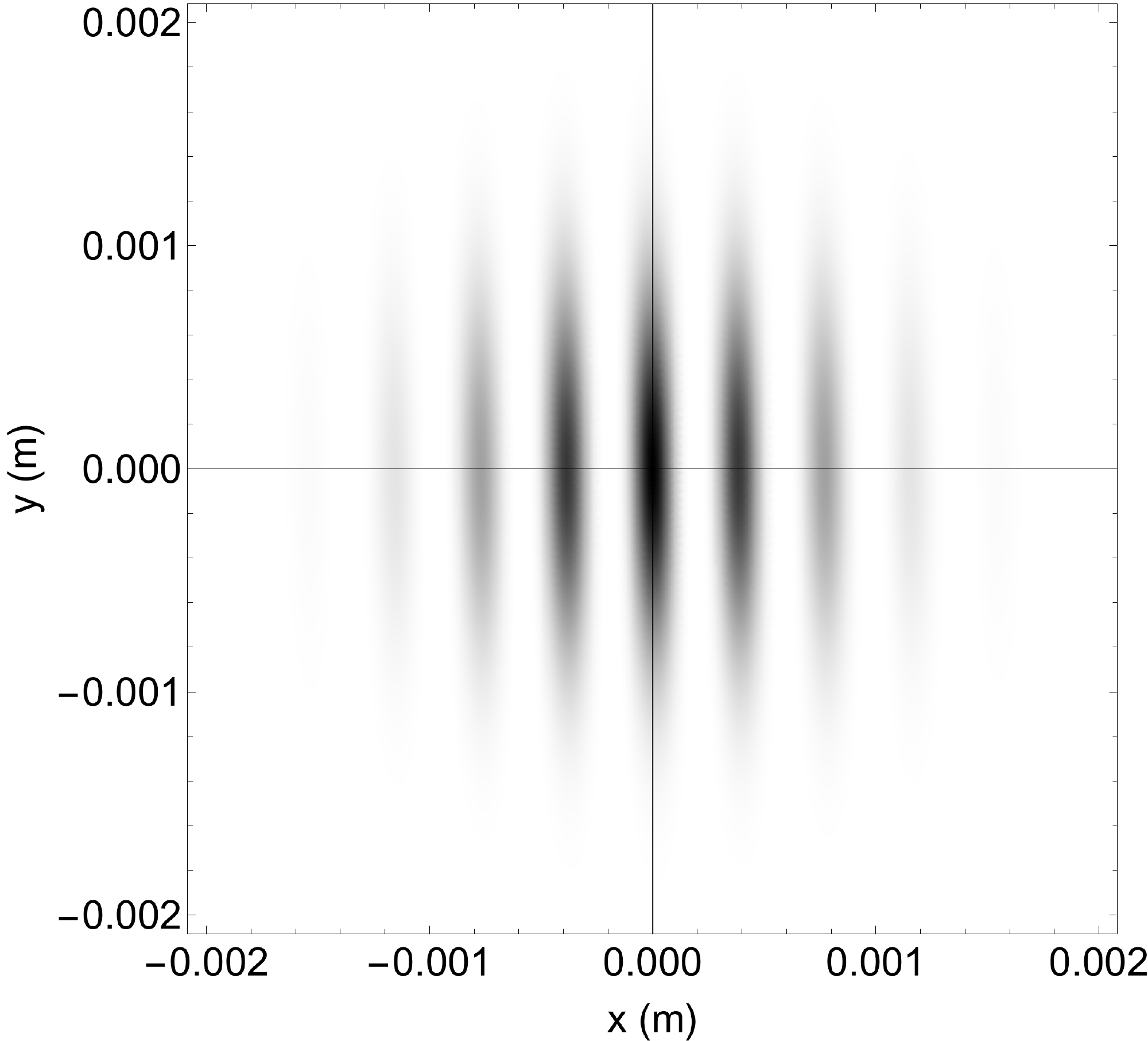}
\includegraphics[width=0.24\textwidth]{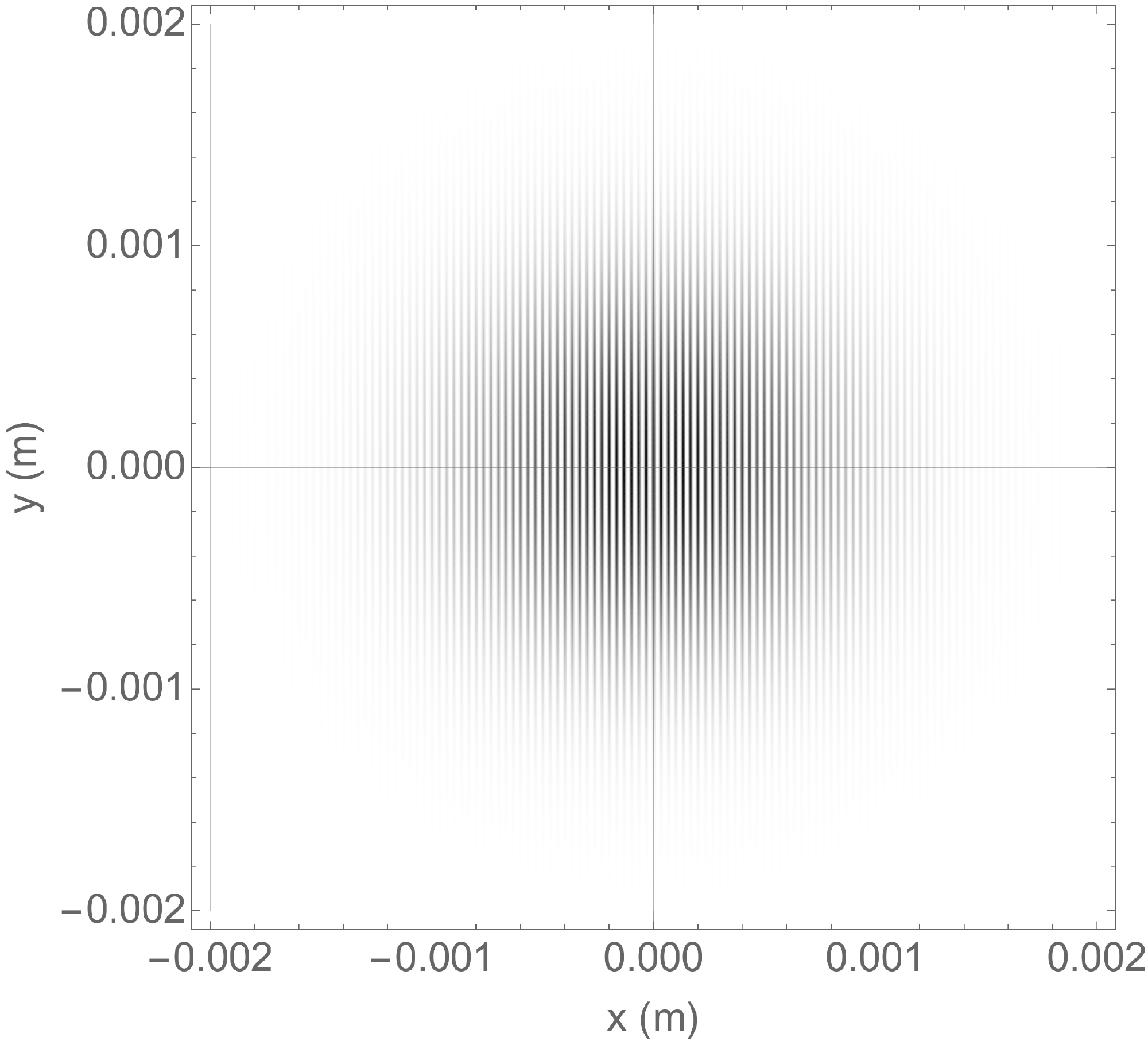}
\caption{\label{fig:fringes} Interferometer fringes obtained for a point-source atomic cloud in the $x$--$y$ plane for different values of $\Omega_y$, from left to the right: $1 \times 10^{-6}$ rad/s,  $1 \times 10^{-5}$ rad/s, $1 \times 10^{-4}$ rad/s, and $1.17 \times 10^{-3}$ rad/s$=\Omega_{\textrm{orb}}$.}
\end{figure}

The $x$--$y$ cross-section of the final density distribution is shown in figure~\ref{fig:fringes} for different values of $\Omega_y$ and when $\Omega_x=\Omega_z=1 \times 10^{-6}$, and $T_{zz}=-2.7 \times 10^{-6}$ s$^{-2}$. The signals are calculated with a bias phase to have the top of a fringe at $(x,y,z)=(0,0,0)$; in the case of a small residual radial velocity, and an interferometer signal spanning over only a fraction of a fringe period, a suitable phase shift can be applied to the interference pattern in order to center the signal at half fringe to increase the phase sensitivity. The increasing angular rotation along the $y$ axis determines an increasing spatial frequency for the fringes along the $x$ axis. The effect of $T_{zz}$ is to spread the phase by $\approx$ 600 mrad along the $z$ direction. When $z$ is chosen as observation direction, the interference pattern encodes the angular velocities along the $x$ and $y$ direction as:
\begin{equation}
\phi(x,y)= 2 k T \left ( \Omega_y x +  \Omega_x y \right ) \, 
\end{equation}
Where the two angular velocities can be obtained with a 2D fit on the atomic fringes. The fringe spacing is inversely proportional to the projection of the angular velocity on the $x$--$y$ plane, i.e. $\Omega_{x-y}=\left ( \Omega_x^2 + \Omega_y^2 \right )^{1/2}$, and it is equal to $\pi/ \left ( k T \Omega_{x-y} \right )$. In order to resolve the interferometer fringes, the CCD camera has to have enough resolution; the requirement becomes more demanding as the satellite angular rotation increases. For example, when $\Omega_{x-y}=\Omega_{\textrm{orb}}$ the fringe spacing is equal to 33 $\mu$m, and imaging 4 mm over 1024 pixels will lead to 3 pixels for each fringe.

\begin{figure}
\begin{minipage}{\textwidth}
  \centering
  \raisebox{0.0\height}{\includegraphics[width=.27\textwidth]{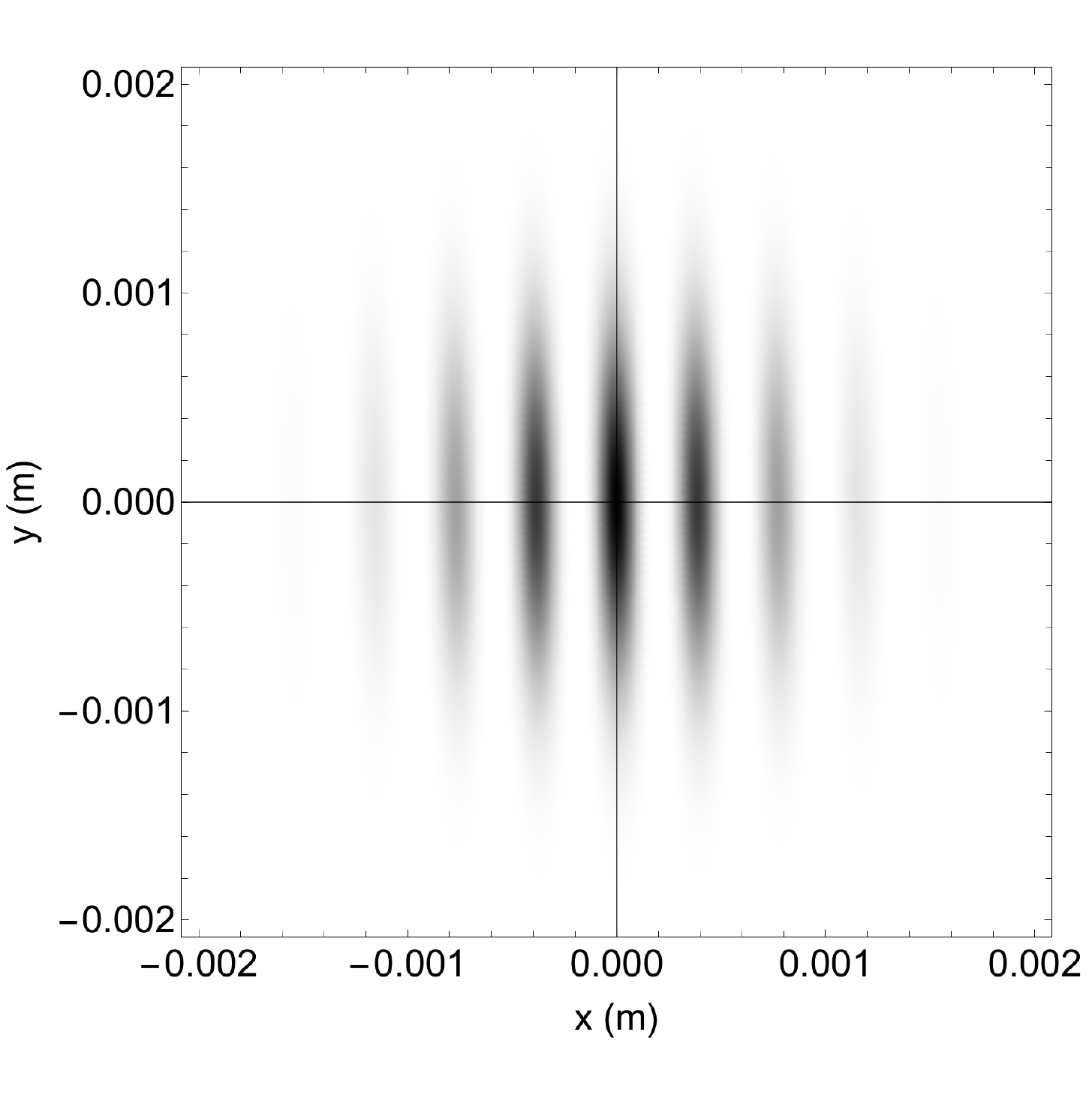}} \hspace*{3mm}
  \raisebox{0.0\height}{\includegraphics[width=.27\textwidth]{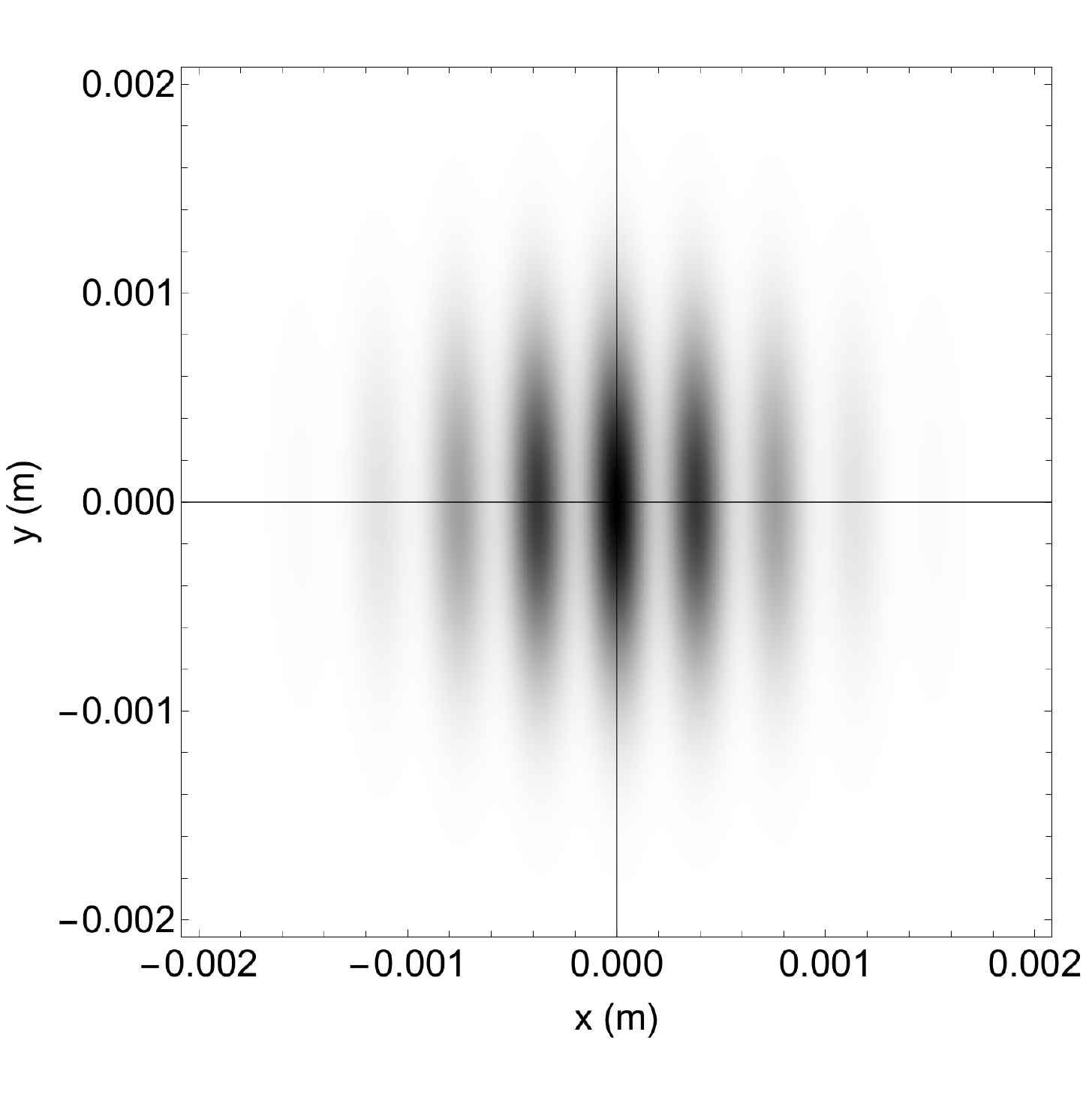}}
  \hspace*{3mm}
  \raisebox{0.13\height}{\includegraphics[width=.37\textwidth]{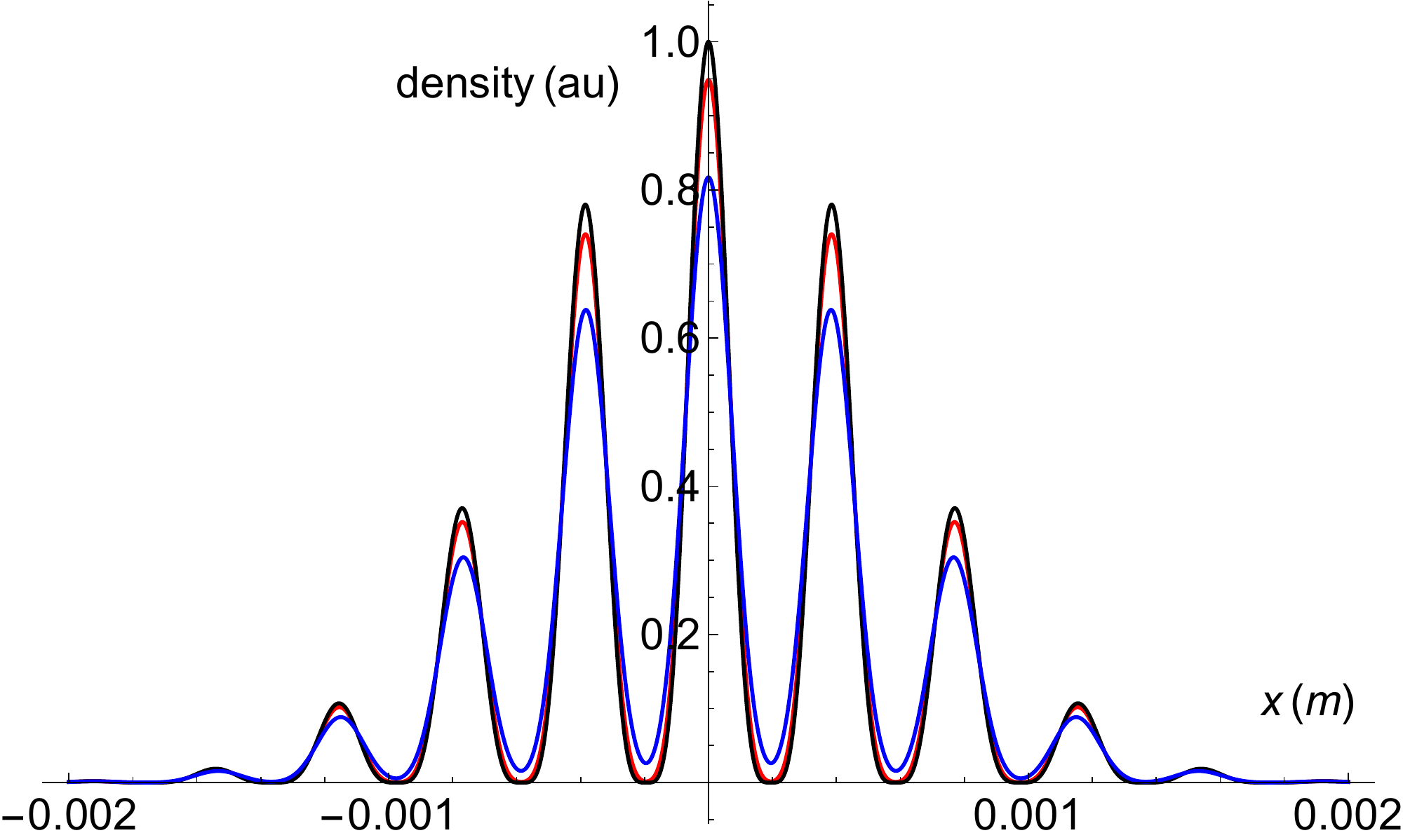}}
  \caption{\label{fig:fringesBlur} (left) fringes on the $x$--$y$ plane at $z=0$ when the initial atomic distribution is taken into account to calculate the final density distribution; a by eye hardly discernible reduction of ~5\% for the fringe amplitude is obtained with respect to the signal resulting from a point-like source, as shown in the third image of figure~\ref{fig:fringes}; (center) fringes on the $x$--$y$ plane when the atomic density distribution is integrated along the measurement direction $z$ for the CCD imaging. (right) The blurring effect on the fringe visibility is shown on the two density distribution profiles taken at $y=0$, when the signal is integrated (red) or not (black) along the $z$ direction. The combined effect is a fringe amplitude reduction of 20\%. The images are obtained for $\Omega_x = 1 \times 10^{-6}$ rad/s, $\Omega_y = 1 \times 10^{-4}$ rad/s, and a final size of the cloud along $z$ of 1.1 mm.}
\end{minipage}
\end{figure}

The initial spatial distribution of the atomic cloud determines the blurring effect on the interferometer fringes; the resulting signal is obtained by calculating the convolution between the probability distribution obtained for the point-like source case with the initial spatial distribution of the atomic ensemble. In figure~\ref{fig:fringesBlur} is shown how the fringes signal worsens when the initial cloud is considered as an isotropic normal distribution along the 3 directions, with a standard deviation equal to 150 $\mu$m. The fringe contrast is further decreased because of the density distribution integration along the observation direction required for the imaging. For $\Omega_y \ll 10$ mrad/s the phase spread along the vertical direction is mainly due to $\Gamma_{zz}$ (see Eq.~\ref{eq:phase_velocity}) and is $\approx$ 600 mrad over the final size of the atomic cloud. Other effects, not taken in to account here, contribute to further reduce the phase sensitivity, like the quantum projection noise (QPN) due to the finite number of atoms detected by each pixel of the CCD, and the technical noise determined by the detection technique \cite{Rocco2014}; these effects must be evaluated to define the requirements for the instrument adopted for the detection.

Note that the above results also apply to the case where the Raman mirrors are tilted to compensate for rotation, provided one replaces $\Omega_y$ with $\Omega_y-\Omega_m$.

\subsection{Design of the laser source/frequency power distribution}

\begin{figure}
  \centering
  \includegraphics[width=16cm]{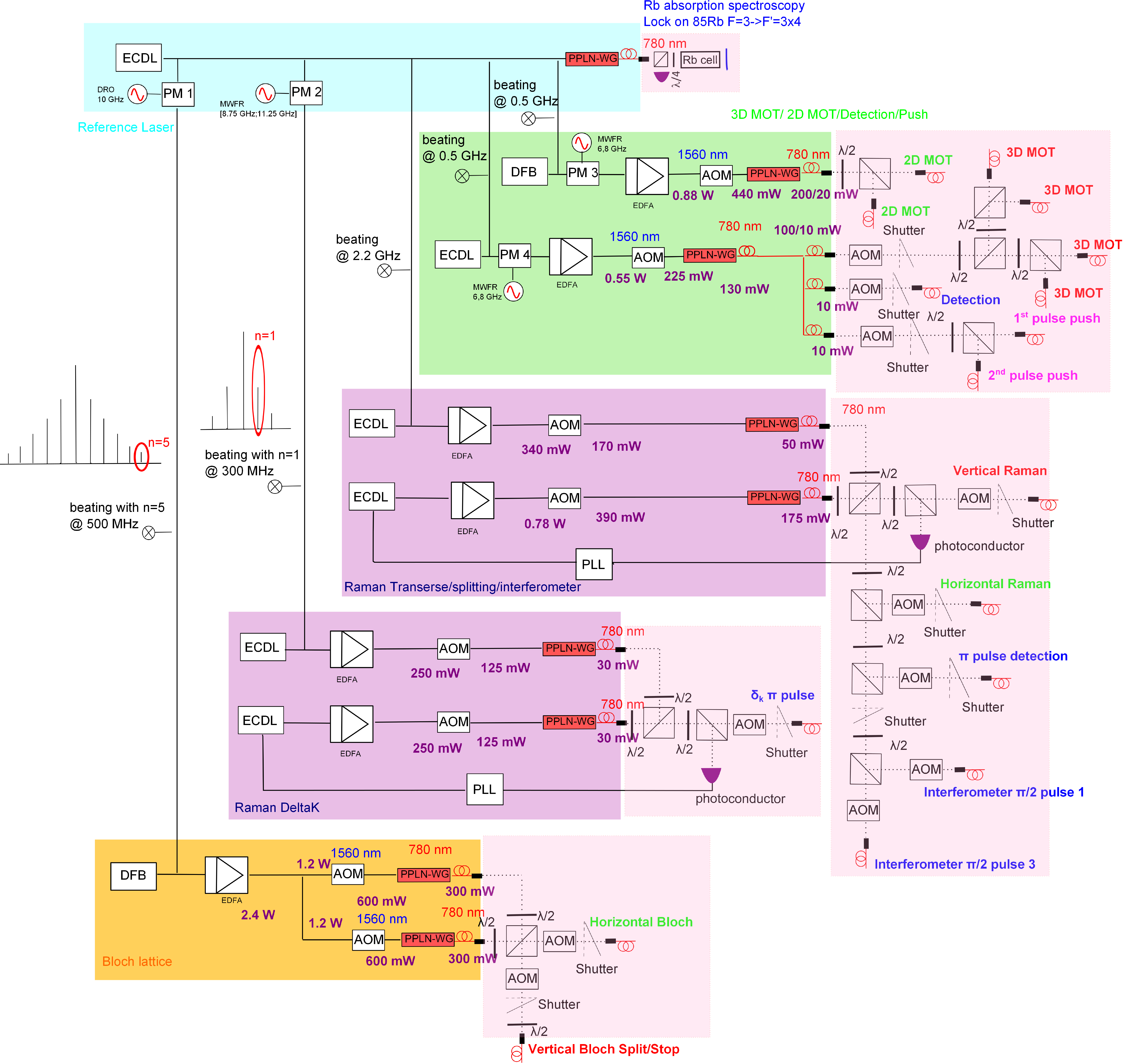}\\
  \caption{Architecture of the laser source. Telecom components (green area for laser cooling, violet areas for the Raman pulses, orange area for Bloch oscillation) are used to develop a compact fiber-based laser system. Free-space optical benches (light pink areas) were implemented with 780 nm optical components to control the frequency of the reference laser (blue area) via saturated absorption spectroscopy and to control the power of the laser output via AOMs. DFB: Distributed Feedback Diode Laser. ECDL: External Cavity Diode Laser. PM: phase modulator. EDFA: erbium-doped fiber amplifier. PPLN-WG: periodically-poled lithium-niobate waveguide. MWFR: microwave frequency reference. DRO: dielectric resonator oscillator. AOM: acousto-optic modulator. FC: fiber coupler. PD: photodiode. PLL: Phase Lock Loop. The needed optical power (violet) are used for the power budget. The frequencies and the power for the $\pi$ pulse laser part on this figure are for $T_{zz}$.}\label{architectureLasersource}
\end{figure}

The architecture of the laser system including the frequency/power distribution is depicted on figure \ref{architectureLasersource}. It is based on telecom technology combined with second harmonic generation (SHG) \cite{Thompson2003}. The main specifications of the laser are detailed in table \ref{laserspecifications}.

\renewcommand{\arraystretch}{1.3}
\begin{table}
  \centering
	\small
  \begin{tabular}{|c|c|ccccccc|}
  \hline
  Specifications & Units& 2D MOT & 3D MOT & Raman & Bloch lattice & Push & Detection &Raman $\Delta k_{z}$\\ \hline
  Wavelength & nm & 780.24& 780.24 & 780.24 & 780.24 & 780.24& 780.24& 780.24\\
  Power & mW &200/20 & 100/10& 45 & 200 & 5& 5&30/30\\
  Frequency & MHz &- 20& [-20;-120]& -3400 & +100000& $\sim 0$& $\sim 0$&[+19100;+24100] \\
  Linewidth & MHz & 0.5 & 0.5 & $<$ 0.01 & $<$ 6& 1& 0.1& $<$ 0.01\\
  Frequency accuracy & kHz &$<$ 100 & $<$ 100 & $<$ 100 & Not critical& $<$ 100& $<$ 100& $<$ 25\\
  Second frequency & NA &F=1$\rightarrow$ F'=2 & F=1$\rightarrow$ F'=2 & 6834+$\omega_{R}$ & [0.075; -3.087] & No &6568& 6834-3 $\omega_{R}$ \\
  Polarization & dB &20 & 20& 30 & 20& 20 & 30&30\\
  Power stability & NA &$< 1\%$ & $< 1\%$ & $< 0.1\%$& $< 1\%$& $< 0.1\%$ & $< 1\%$& $< 0.1\%$\\
  Disruptive lines & NA &Tolerated & Tolerated & No & No& No& No&No\\ \hline
  \end{tabular}
  \caption{Main specifications of the laser source for $T_{zz}$. Each specification is determined for one type of function. For instance the ``Raman" column is for both the atom motion and the $\pi/2$ interferometer pulses since the specifications are identical. The required power is for a single Raman beam. Since they are used simultaneously, this value need to be multiplied by the number of beams in the total power budget. The two values for the laser power correspond respectively to one optical frequency. Disruptive lines can be produced if we use a phase modulator to create the second optical frequency.  $\omega_{R}$ is the recoil frequency. The difference about the laser parameters for $T_{xx}$ and $T_{yy}$ are the frequency detuning range ([-17150;-14650] MHz) and the laser power (22/22 mW) of the interferometer $\pi$ pulse. }\label{laserspecifications}
\end{table}
\renewcommand{\arraystretch}{1.0}

A reference laser supplies an absolute optical frequency corresponding to the atom transition of Rubidium ($^{85}$Rb crossover transition $|F=3\rangle\rightarrow|F'=3(4)\rangle$). Then all the laser frequencies are servo locked compared to this reference thanks to a beatnote and a lock-in electronics box. The laser system is composed of four other blocks corresponding to the functions of Table \ref{laserspecifications}: one block for the cooling and the detection, two subsystems for the Raman transitions (one for the $\pi$ pulse of the interferometer, one for the other pulses), and the last part corresponding to the implementation of the Bloch lattice. The wave vector modification of the $\pi$ pulse interferometer $\delta k$ implies an additional laser system. Indeed, the laser system to generate $\pi$ and $\pi/2$ pulses of the atom interferometer cannot be the same because the frequency difference is too large and they are used simultaneously. The reference frequency for the $\pi$ Raman pulse is produced using a phase modulator (PM 2) with a tunable Microwave Frequency Reference (MWFR). The reference frequency for the Bloch lattice laser is the fifth harmonics at +50 GHz of a frequency comb generated using a phase modulator (PM 1) with a DRO operating at 10 GHz.

The laser source for the cooling part is a standard DFB diode.
A phase modulator (PM 3) with a DRO at 6.8 GHz creates the repumping frequency used for laser cooling in the 2D MOT. This beam is then amplified by an erbium-doped fiber amplifier (EDFA). Frequency doubling from 1560 nm to 780 nm is accomplished via SHG in a periodically-poled lithium niobate waveguide (PPLN-WG), where the confinement of the optical mode leads to high intensity and thus high efficiency. A similar architecture is implemented for the 3D MOT, the detection and the push beam except for the laser diode (External Cavity Diode Laser). A fibered splitter is used to separate the beams for the three functions.

The configuration of the Raman laser source is based on the generation of the two frequencies by distinct external cavity diodes (ECDL). The two lasers are phase locked thanks to a beat note on a photoconductor at the output of the laser system to avoid phase noise along the optical paths. We duplicate the Raman laser system for the interferometer $\pi$ pulse which has a specific detuning and power compared to the $\pi/2$ pulses due to the shift of the wave vector $\delta_{k}$. The required frequency difference between the $\pi$ pulse and the $\pi/2$ pulses to compensate for $T_{zz}$ is +25 GHz. Since the gravity gradient is not constant during the full orbit and the full mission, this frequency shift has to be tunable over a range of $\pm 2.5$ GHz. Moreover, the idea is to actively compensate for the phase shift due to the variation of the gravity gradient, and extracting the measurement from the needed frequency shift. This requires a control of the frequency shift with a relative accuracy of $10^{-6}$ to be compliant with the specification of 5 mE per shot for the gravity gradient measurement.

The laser source for the Bloch lattice subsystem is a DFB diode. To create the two frequencies, the beam is split into two paths and a different RF frequency is supplied on each fibered acousto-optic modulators (AOMs). The two beams are then recombined on the free space combination bench.

 ZERODUR\textregistered free-space platforms hold AOMs and mechanical shutters to switch on and off the output light. An assembly of polarizing cubes and waveplates splits and recombines the beams.
 Two laser sources with optical frequencies separated by the clock transition of Rubidium 6.8GHz are combined for the Raman pulses. The polarization are crossed and the two output of the polarizing cubes are exploited in order to avoid power losses. A beat-note is done on a photoconductor and the second laser source is phase locked on the first one.

\subsection{Reference mirror design}

The mirrors discussed in this section retroreflect the beamsplitting light fields. They serve as a reference for the effective wave fronts of the light fields used to coherently manipulate the atoms. Ideally, these effective wave fronts are flat and smooth over all atom-light interaction zones to avoid spurious phase contributions. This condition implies a parallel alignment of the surfaces in the cross-track direction, and well aligned tilts in the other two directions which compensate for rotations~\cite{Freier16CS,Dickerson2013PRL,Lan2012PRL} and lead to effective parallel wave fronts seen by the atoms. The retro reflection setup suppresses inhomogeneity of the incoming light field, but imperfections in the ideally flat mirror surface directly affect the effective wave front. Wave front distortions are a major contribution to the uncertainty in precision atom interferometers~\cite{Freier16CS,Gauguet2009PRA,Louchet2011NJP,Berg2015PRL}; this noise can be reduced using colder atoms~\cite{Louchet2011NJP,Karcher2018NJPL}, and further minimized with condensed~\cite{Rudolph2015} or even collimated atoms~\cite{Abend2016PRL,Kovachy2015PRL,Muntinga2013}. These considerations motivate the detailed discussion for the gradiometer. We assess the requirements on the mirror in three steps: (i) tilts, (ii) defocus and Gouy phase, (iii) higher order distortions, and propose a technical implementation. The following assessments treat a Gaussian distribution with an initial standard deviation of the position of $400\,\mu\mathrm{m}$, and of the velocity of $100\,\mu\mathrm{m/s}$ for an atomic ensemble of $10^6$ atoms.

For assessing the impact of relative tilts of the effective wave fronts, we follow the description of Ref.~\cite{Tackmann2012NJP}. To first order, and for small angles $\alpha$, a tilt implies a change in the distance of the atoms to the retro reflection mirror of $d\cdot\sin{\alpha}$. Here, $d$ denotes the distance between the pivot point and the projection of the position of the atoms onto the mirror surface in direction of the beam splitting light field. Although the choice of $d$ has a strong impact onto the phase of a single atom interferometer, it is suppressed in the differential signal. This suppression depends on the overlap of the trajectories parallel to the mirror surfaces~\cite{Fils2005EPJD} which itself depends on the precision of the launching mechanism, that transfers the atoms into the interferometer region, and ultimately on the shot-noise limited position and velocity uncertainty of the atomic ensembles. For our gradiometer, we have to control phase noise, and can neglect an unknown, but stable phase shift.
\begin{figure}[tp]
\centering
\includegraphics[width=10cm]{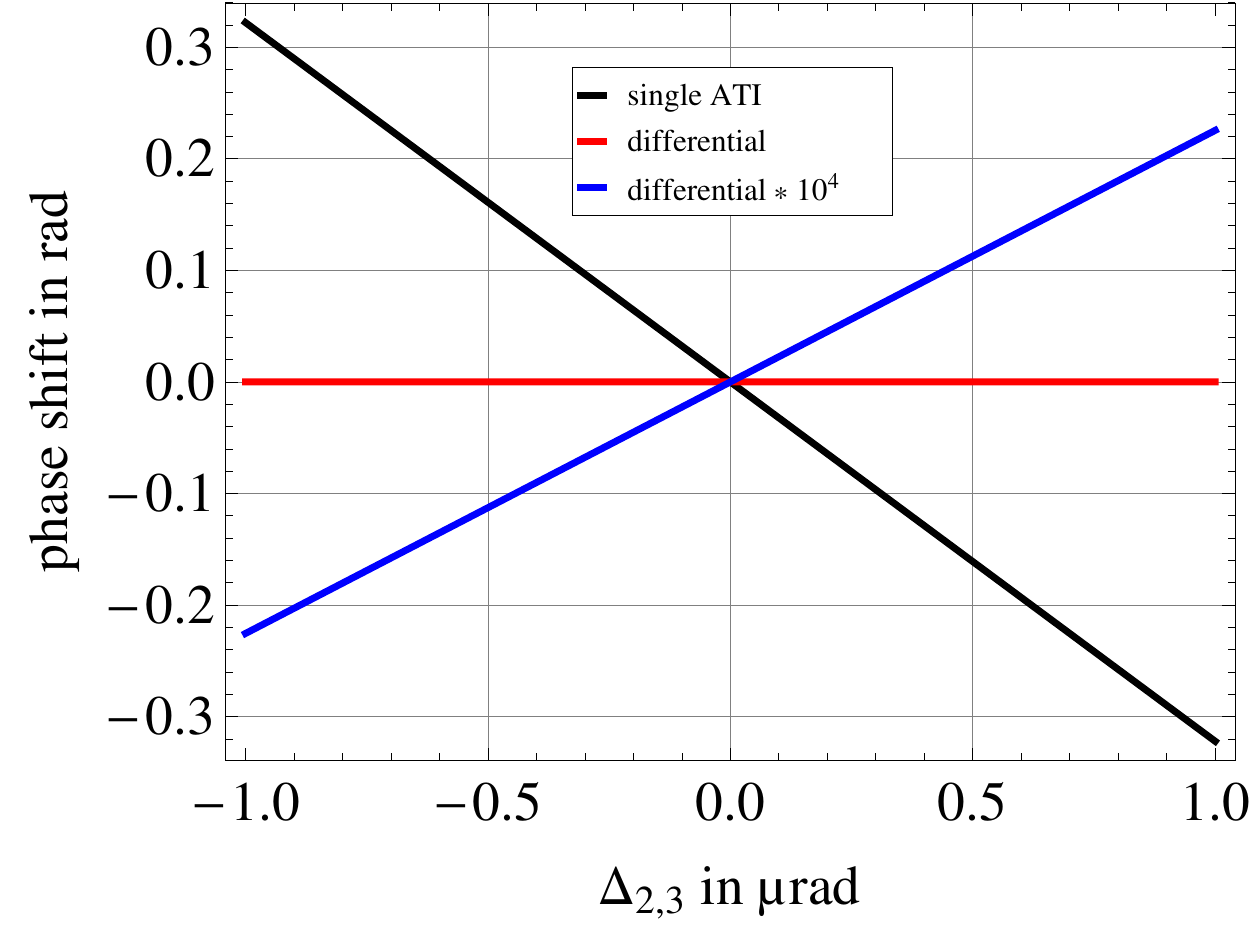}
\caption{Impact on the phase of a relative tilt of the wave fronts.
The figure shows the case for a relative tilt $\Delta_{2,3}$ of the wave fronts in direction of the forward drift velocity in the second and third interaction zone. While a significant impact on the single atom interferometer (black) is visible, the effect is strongly suppressed in the differential phase (red).
The blue curve exaggerates the phase by a factor of $10^4$ for clarity.
The perpendicular direction shows the same behaviour.
}
\label{fig:tilt}
\end{figure}
Figure~\ref{fig:tilt} shows an example of a phase shift induced by the relative tilt between two effective wave fronts and its suppression in the differential signal when assuming a shot noise limited differential starting position and velocity (cross-track direction). From the slope of the curves we derive the requirement to keep relative tilts below $25\,\mu$rad [$0.5\,\mu$rad] if the relative position jitter between two subsequent cycles is below $2\,\mu$m [$100\,\mu$m], which is at or above the shot-noise limit ($<2\,\mu$m). This corresponds to a peak to valley of the mirror surface of 250 nm [5 nm] over a region with a diameter of 1 cm. The same requirements are valid for the other two axes (along track and nadir) when interpreting the relative mirror tilts as the deviation from the ideal alignment for compensating the rotations.

In a simplified model~\cite{Louchet2011NJP}, an acceleration signal is contaminated by $\delta\,a=\sigma_v^2/R$ for an expansion rate $\sigma_v$ of the atomic ensemble, assuming a Gaussian distribution, and a radius $R$ of a static effective wave front curvature.
In our more detailed model the wave front curvature changes as the beam propagates, and the effective wave front is generated by subtracting the retro reflected beam from the incoming one~\cite{Schubert2013arXiv}. We propagate the beam parameters via ABCD matrices~\cite{Kogelnik1966ApplOpt}, determining the wave front curvature and Gouy phase depending on the longitudinal position inside the light field. The wave front of the incoming beam depends on the numerical aperture of the fiber, a free propagation, the focal length of the lens for collimating the beam, and another free propagation to the atoms. The wave front of the retro reflected beam depends on the numerical aperture of the fiber, a free propagation, the focal length of the lens for collimating the beam, a free propagation to the mirror (past the atoms), the focal length of the lens describing the mirror, and a free propagation back to the atoms. We take the different positions of the atoms on the upper and lower trajectories of the interferometer into account, and calculate the phase for both interferometers forming the gradiometer. Our model also enables the evaluation of cases where the atoms are off center with respect to the beam. In addition, it introduces a weighting according to the Gaussian intensity profile, since transition probabilities at the edge of the beam degrade because of reduced Rabi frequencies, and atoms at these positions contribute less to the average phase.

\begin{figure}[tp]
\centering
\includegraphics[width=10cm]{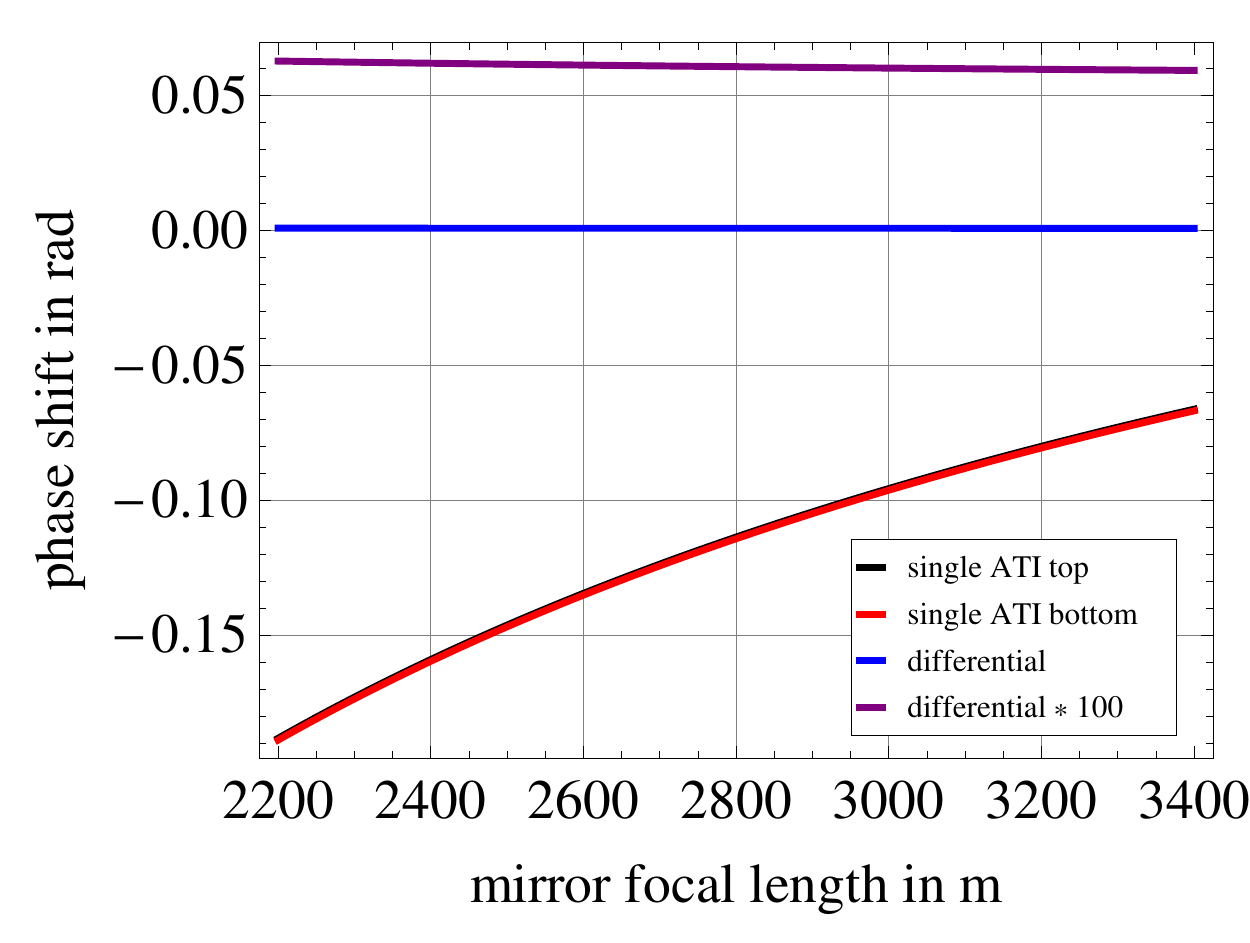}
\caption{Impact on the phase for different focal lengths of the mirror.
The figure shows the case for an approximate waist of 10\,mm, third interaction zone for recombination, in the tilted-mirror configuration.
As can be seen, the differential phase is robust against variations of the focal length within the given parameters, and significantly suppressed compared to a single atom interferometer (single ATI top / bottom). To better illustrate the phase behavior, the blue signal has been multiplied by 100 and reported in purple. The other two zones have a qualitatively similar behaviour.
}
\label{fig:defocus}
\end{figure}

We initially randomize the values for longitudinal positions of the optics in the beam (standard deviation of 5 $\mu$m to 20 $\mu$m), the focal length of the collimation lens (standard deviation of few 10 $\mu$m), the velocities and positions of the atomic ensembles (shot noise limited standard deviation of 0.1  $\mu$m/s, and 0.4 $\mu$m, respectively), and the curvature of the lens modelling the mirror (standard deviation of 300 m) to slightly deviate from the ideal value. For the mirror, we assume a mean curvature radius of $R$=5600 m. This corresponds to a peak to valley of $\lambda$/20 for $\lambda$=780 nm and diameter of the mirror of 6 cm and leads to a reflected wave front with a curvature radius of $R$=2800 m. We individually scan the focal lengths of the collimation lens and the mirror for each interaction zone and keep other parameters fixed to evaluate the impact on the phase of the interferometer for approximate beam waists of 2.5 mm, 5 mm, and 10 mm. An example curve is shown in figure~\ref{fig:defocus}.
In addition, we take the standard deviation of the phase for 1000 shots with randomized values for the optics positions in longitudinal direction (standard deviation of few $\mu$m), the positions of the beam splitting axes in transverse direction (standard deviation 100 $\mu$m), and the velocities and positions of the atomic ensembles (shot noise limited).

For the cross-track direction, with an approximate waist of 2.5 mm [5 mm, 10 mm], we estimate a phase noise per shot in the gradiometer signal of about 1 mrad [0.2 mrad, 0.05 mrad] and a phase bias of about 4 mrad [4 mrad, 1 mrad] or less for a deviation of the focal length of the collimation lens from the optimum of $\leq5\,\mu$m [40 $\mu$m, 40 $\mu$m] or less and a deviation of the focal length of the mirror of $\leq$600 m. For waists above 2.5 mm, the Gouy phase dominates the phase excursion in a single atom interferometer. Although it still dominates the residual phase shift in the differential signal for the gradiometer in our simulation, it is sufficiently suppressed.

For simulating the other two axes, we add displacements to the positions of the atoms with respect to the beam center as required by the tilted-mirror configuration. Again using an approximate waist of 2.5 mm [5 mm, 10 mm], we estimate a phase noise per shot in the gradiometer signal of about 3 mrad [1 mrad, 3 mrad] and a phase bias of about 10 mrad [5 mrad, 1 mrad] or less for a deviation of the focal length of the collimation lens from the optimum of $\leq5\,\mu$m [40 $\mu$m, 40 $\mu$m] or less and a deviation of the focal length of the mirror of $\leq$600 m. Here, the quality of the input collimation dominates the residual shift for waists of 2.5 mm and 5 mm, and for 10 mm the dependence on the Gouy phase is qualitatively similar to the previous case. To push the noise of the configuration with a waist of 10 mm below 1 mrad, the standard deviation of the positions of the beam splitting axes in transverse direction has to be limited to 25 $\mu$m.

Consequently, this validates our choice of a mirror with a curvature radius of about $R$=5600 m, and sets the requirement for minimum beam waist of 5 mm.

To simulate higher order distortions, we follow the approach of Refs.~\cite{Fils2005EPJD,Gauguet2009PRA,Tackmann2009Dipl} by parameterizing the atomic trajectories and calculating the local phase shifts imprinted by the atom-light interactions which are affected by imperfect mirror surfaces, represented by reference objects. Here, we do not propagate the wave fronts, and simply assume the effective wave fronts to be a copy of the surface inhomogeneities. The three reference objects have a peak-to-valley figure of about $\lambda$/20 [$\lambda$/75] and a root mean square of $\sim\lambda$/100 [$\lambda$/150] over the whole surface [central region, diameter 1 cm]. Due to reflection, these values are doubled in the simulation. We take the standard deviation of 10 averages with 400 shots each for randomized positions [velocities] of an atom with a standard deviation of 400 $\mu$m [100 $\mu$m/s]. This procedure is repeated for four different center positions of the reference objects and two different sequences of the reference objects to rule out readings from particularly good or bad spots. The simulation returns an estimate for the differential phase noise of 10 mrad per shot, consistent between the individual intializations. This is a factor of 10 above the requirements. Since the results indicate uncorrelated noise between the two single interferometers contributing to the gradiometer signal, we expect these results to hold for the tilted mirror configuration.

The simulations imply the requirements of having a local peak-to-valley figure of $\leq\lambda$/1000, and a local root mean square of $\leq\lambda$/1000 ($\sim$1 nm). For designing appropriate optics, the simulation can be adapted by replacing the reference objects with two blanks and an object whose surface is defined by a single Zernike polynomial~\cite{Prata1989}, and determining the required pre-factors.

Summarizing these assessments, the requirements are a peak-to-valley figure of $\lambda$/20 for a mirror with a diameter of 6 cm, a peak-to-valley figure of $\lambda$/1000 ($\sim$1 nm) in the central region with a diameter of 1 cm, a root mean square of $\lambda$/1000 ($\sim$1 nm) in the central region with a diameter of 1 cm, and a maximum relative tilt of the effective wave fronts in the beam splitting zones below 0.5 $\mu$m. When assuming a fused silica substrate for the mirrors, temperature gradients have to be limited below 63 K/m to avoid distortions which violate the requirements above.

Mirrors in gravitational wave detectors as VIRGO~\cite{Beauville2004CQG,Tesar1992ApplOpt} and LIGO~\cite{Granata2016PRD,Acernese2015CQG} are based on fused silica substrates coated by ion beam sputtering with SiO$_2$ and Ta$_2$O$_5$ or TiO$_2$. Here, the surface quality of the substrate dominates the inhomogeneity after coating. Test substrates with a diameter of 48 mm and a thickness of 12 mm were polished to a peak-to-valley figure below 1 nm and a root mean square of 0.2 nm~\cite{Middleton2006OptEng,Antonucci2011CQG,Harry2010CQG}. In LIGO, a test sample of a coated substrate reached a roughness below 0.5 nm root mean square, and the requirement on controlling the curvature of the mirror at the level of $\pm$2 m implies a control of about 1 nm in the peak-to-valley figure.

Since these values are compatible with our requirements, we propose to use the same mirror technology for the gradiometer. For the cross-track axis we propose a single mirror covering all three atom-light interaction zones for passive stability with a size of 6$\, \times \,$6$\, \times \,$31 cm$^3$, leading to a mass of 2.5 kg, and for the other two axes three separate mirrors with a diameter of 6 cm and a thickness of 1.5 cm. The mirrors should be mounted inside the vacuum system to avoid additional distortions by viewports. In the axes with tilted mirror configuration, the outer ones require motorized mirror mounts for initial alignment and adjustment to the actual rotation rate.

\subsection{Magnetic field}

\subsubsection{Requirements} \label{Sec:MagFieldReq}

For the interferometer region a static bias field of e.g. $B$=100 nT aligned with the Raman lasers is applied to provide a quantization axis for the atoms. External magnetic fields need to be attenuated below this level by passive shielding and temporal and spatial fluctuations of the magnetic fields need to be suppressed since they cause phase shifts due to the second order Zeeman effect and bias in the measurement of the gravity gradient. 

With a gradient in the magnetic field $\nabla B$ and the quadratic Zeeman shift in energy $\delta E=hKB^2$ with $K$=575 Hz/G, the atoms accelerate by $\delta a = \frac{-2hK}{m} B\nabla B$. Given the same constant spatial gradient in both interferometers, the differential acceleration between the two interferometers then is $\Delta(\delta a)=\frac{-2hK}{m}(\nabla B)^2 \Delta z$. This results in a bias on the gravity tensor 
\begin{equation}
\Delta T_{zz}=\frac{\Delta (\delta a)}{\Delta z} = \frac{-2hK}{m}(\nabla B)^2.
\end{equation}
In order to reduce the bias on the gradiometer phase below 1 mrad, the magnetic field gradient needs to be reduced below 60 nT/m.

Similarly, a difference of the time averaged field $\langle B \, \rangle$ between the first and second half of the pulse sequence leads to a bias in the phase measurement of $\Delta \phi = 4 \pi K B \Delta \langle B \, \rangle T$. Thus with $B$=100 nT a field fluctuation by $\Delta \langle B \, \rangle$=14 nT causes a phase shift of 1 mrad. This is then subject to common mode suppression in a differential measurement depending on the spatial and temporal correlations of the fields at the two interferometers. 

All of the above requirements need to be met, as the satellite moves in Earth's magnetic field of $B_{\rm{earth}} \approx 40~\mu$T, causing a change of the field component projected on the interferometer axis of up to 60 nT/s. Thus, an efficient mumetal magnetic shield is required to provide a suitable magnetic field environment \cite{Dickerson2012}.

\subsubsection{Magnetic shield design} \label{Sec:MagShieldDesign}

To provide shielding of the interferometer against external magnetic fields, we consider a passive multilayer mumetal shield. The effectiveness of the magnetic shield can be described by the shielding factor
\begin{equation}
S = \frac{B_{outside}}{B_{inside}},
\end{equation}
defined by the ratio of the total residual magnetic field inside the shield to the total initial magnetic field outside. The magnetic field requirements derived above thus translate into a required shielding factor of $S>1000$.

\begin{figure}[ht]
\centering	
\includegraphics[width=0.35\textwidth]{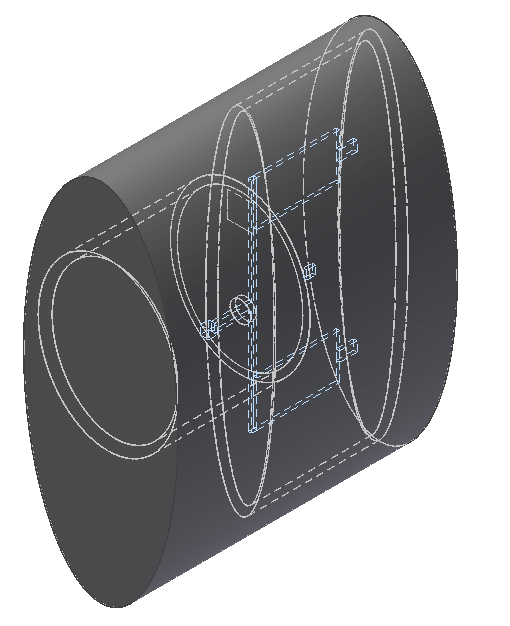}
\caption{Magnetic shield design, consisting of one 1 mm outer layer and two 1 mm inner layers for each vacuum chamber. The blue dotted area depicts the location of the atoms inside the shield.}
\label{fig:MagShieldDesign}
\end{figure}

To verify that our design complies with this requirement, we performed simulations based on finite-element modeling (FEM), similar to \cite{MuShield2016}. The modeled mumetal shield was placed inside a static magnetic field $B_{\rm{earth}}$=40 $\mu$T and the residual field at the interferometer region was calculated to determine the shielding factor and the resulting magnetic field gradient along the interferometer area. Starting from a cylindrical design the best trade-off between shielding effectiveness and dimension and mass was found with elliptical shields. The 3-layer magnetic shield design, shown in figure~\ref{fig:MagShieldDesign}, consists of one overall outer layer and two inner layers for the 2D-MOT/BEC chamber and the CAI chamber respectively. The gaps between the layers were set to 20 mm and each layer has a thickness of 1 mm.

The FEM simulation results in a shielding factor of $S_{x,y,z}\approx(49\,700, 9\,900, 4\,800$) for the interferometer area which complies with the requirement. The actually achievable shielding factor will of course depend on the fabrication and details of the design such as location of necessary feedthroughs which are not yet included here. Thus, we take these results only to confirm that a three-layer shield should be capable to meet the requirement in principle. For the magnetic gradient we observe a maximum value of $\Delta B_{x,y,z}=(0.74, 2.7, 4.6)$ nT/m, also fulfilling the requirement. These values are determined from the maximum observed differences of the total residual fields over the two interferometer regions and thus represent an upper limit to the gradient.

\subsection{Payload architecture}

\begin{figure}[ht]
\centering	
\includegraphics[width=0.8\textwidth]{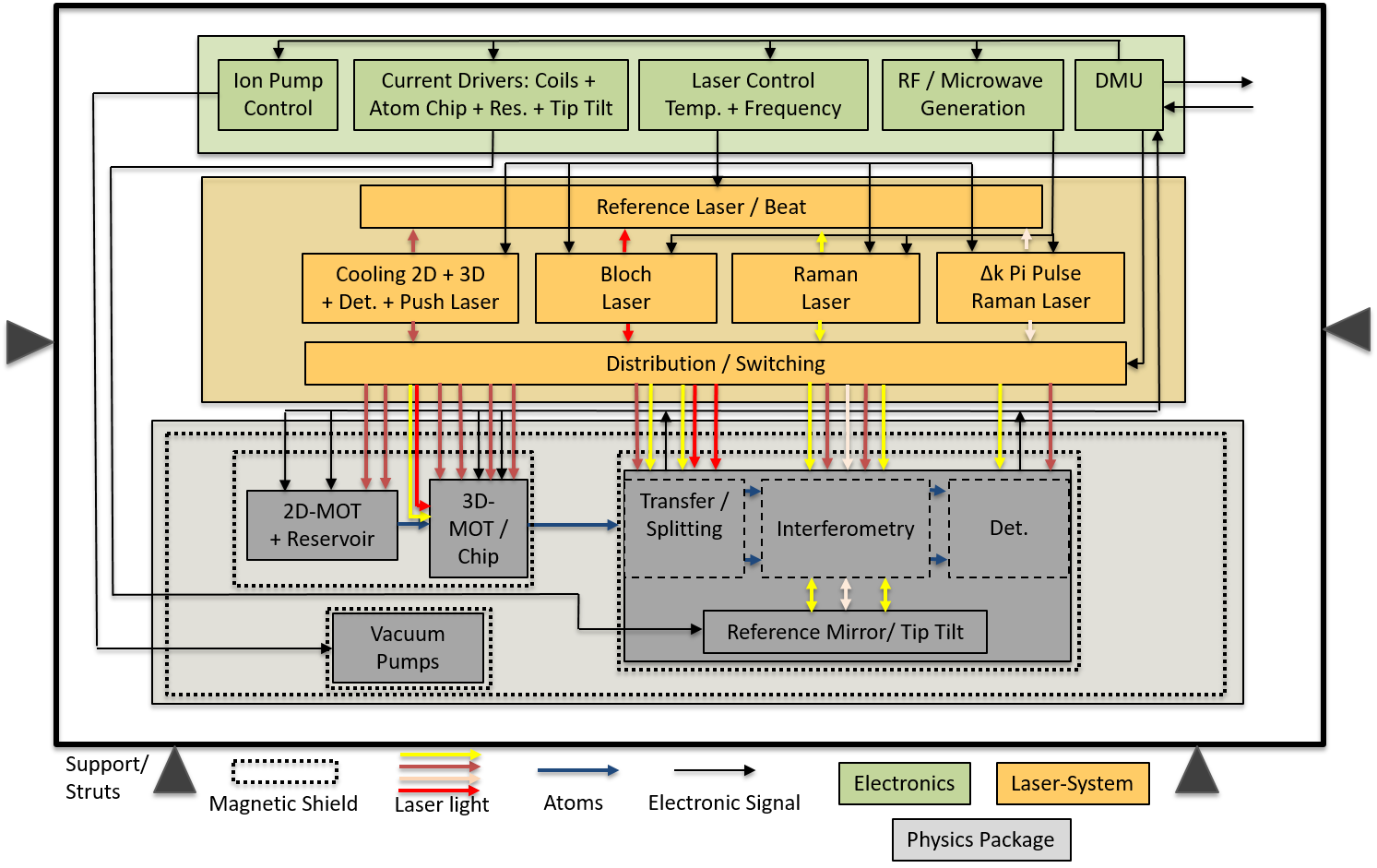}
\caption{Functional diagram of a single axis gradiometer instrument in CAI.}
\label{fig:FuncDiag}
\end{figure}

Here we consider a satellite in nadir-pointing configuration with a payload of three differently orientated instruments to measure the gravity gradient in each spatial direction: $T_{xx}$, $T_{yy}$ and $T_{zz}$. Each of the three instruments is divided into three functional units: the physics package, the laser system and an electronics unit as depicted in the functional diagram of figure~\ref{fig:FuncDiag}. In table \ref{Tab:zOverview}, we give a breakdown of the mass and power budget for the single $T_{zz}$ instrument into these three units. The mass and power for the $T_{xx}$ and $T_{yy}$ sensors slightly differs due to the additionally required tip-tilt mirror. This budget is derived from the design presented in section \ref{Sec:Design} and based on state of the art technology of cold atom experiments in microgravity \cite{Rudolph2015,Muntinga2013}. 

\begin{table}[ht]	
	\centering
	\begin{tabular}{|c|c|c|c|}
		\hline
	System & Mass (kg) & Power  (W) & Size (mm)\\
		\hhline{====}
		Physics Package & 160.4 & 128.4 & (1052 x 444 x 805)\\
		\hline
		Laser System & 51.1 & 104.2 & (300 x 300 x 400)\\
		\hline
		Electronic System & 52.4 & 607.9 & (300 x 300 x 1000) \\
		\hhline{====}
		{\bf Total } & \bf 264.0 & \bf 840.9 & (1052 x 750 x 805) \\  		
		\hline
	\end{tabular}
	\caption{Mass and power budget for a $T_{zz}$ instrument including the tip-tilt mirror design. A component margin of 20\% on mass and power is included here.}
	\label{Tab:zOverview}
\end{table}

From the basic vacuum chamber design in figure~\ref{VacuumSchematics}, we estimate the size of the physics package for a single axis instrument to be (1052$\, \times \,$444$\, \times \,$805) mm$^3$ including the elliptical magnetic shield design, discussed in section \ref{Sec:MagShieldDesign}. An arrangement of all three instruments including lasers and electronics then results in an approximate estimate of the payload size of (1054$\, \times \,$1054$\, \times \,$1600) mm$^3$. This allows for an elongated  satellite shape similar to that of GOCE with a front surface slightly larger than 1 m$^2$, which is important to minimize the residual atmospheric drag of the satellite. The power and mass of a full 3-axis instrument including a $20\%$ margin is estimated at 785 kg and 2940 W. However we expect that ongoing technology development should allow for a significant reduction of these values.

\section{Performance analysis and optimum measurement bandwidth}

\label{Performance_analysis}

In this section, we quantify precisely how the CAI gravity gradiometer can improve our knowledge on the Earth's gravity field. As such, we designed a closed-loop numerical simulator and applied it to evaluate the performance of the CAI gradiometer. We focused our analysis on the nadir pointing mode with the compensation of rotation. 

\subsection{Closed-loop simulator}
\label{Simulator}

In order to represent the chain of the measurement process realistically, we carried out this study through a closed-loop simulation in the time domain. This approach is flexible and provides the possibility to precisely quantify the errors in terms of gravity field solutions.  

\begin{figure}[!htb]
	\centering
	\includegraphics[width=0.8\textwidth]{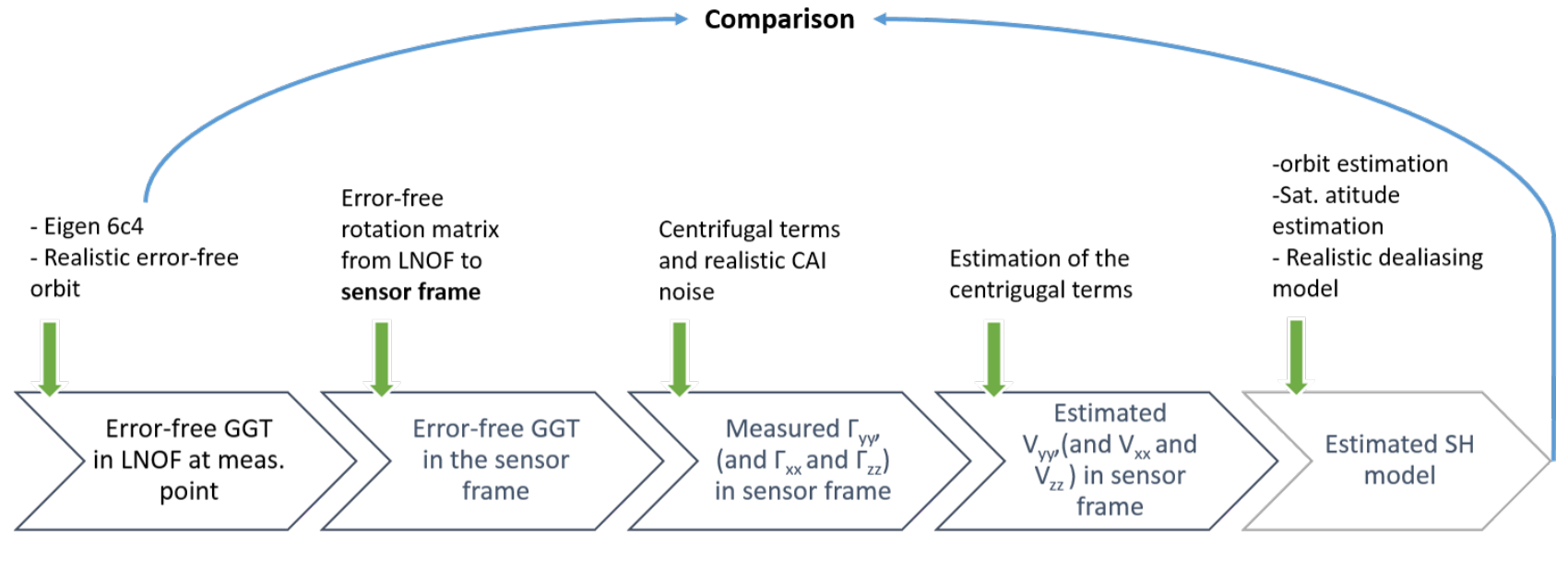}
	\caption{Workflow diagram of the simulation principle, focused on the processing of the gravity gradients.}
	\label{Fig_Workflow}
\end{figure}

A general overview of the closed-loop simulation workflow is given in figure ~\ref{Fig_Workflow}. On the one hand, the simulator takes a gravitational model and the noise-free time series of the satellite orbit, angular velocity and attitude, and, on the other hand, the spectral or statistic characteristics of the sensors' noise, to synthesize realistically degraded observables, namely the estimated orbit, angular velocity, attitude and gradients. These synthesized observables are finally used to derive a gravity field model. The comparison of the estimated gravity field model with the original one enables then to precisely characterize the error in the frequency and spatial domain. 

\subsection{Gravity field recovery}
\label{Recovery}

The Earth's global gravity field is usually expressed in terms of a Spherical Harmonics (SH) series \cite{Hofmann-Wellenhof2006} as 
\begin{equation}
	\label{eq_gravity_field_model}
	V (r,\theta,\lambda) = \frac{GM}{R} \sum_{n=0}^{N} (\frac{R}{r})^{n+1} 
	\sum_{m=0}^{n} \left[ \bar{C}_{nm} \cos m\lambda + \bar{S}_{nm} \sin m\lambda \right] \bar{P}_{nm}(\cos \theta),  
\end{equation}
where $GM$ is the gravitational constant of the Earth, $R$ is the radius of Earth, $(r,\theta,\lambda)$ are spherical coordinates of a point on the Earth surface ($r$ radius, $\theta$ co-latitude, $\lambda$ longitude), $n,m$ denote SH degree and order, $N$ denotes the maximum degree of the model expansion (in theory, the maximum degree is $\infty$), $\bar{P}_{nm} (\cos \theta)$ are fully normalized associated Legendre functions, and $\bar{C}_{nm}, \bar{S}_{nm}$ are the normalized SH coefficients, which are the unknowns of the gravity field solution. 

The gravity gradients $V_{ij}$ are point-wise measurements of the second-order partial derivatives of the gravity potential. They are usually delivered in the instrumental reference frame, e.g., the Gradiometer Reference Frame (GRF), while the Earth's gravity field model is expressed in the Earth-fixed Reference Frame (ERF). We thus have to transform the gravity gradients and the gravity field model to the same reference frame by
\begin{equation}
	\label{eq_Func_model}
	V_{ij} = R \frac{\partial^2 V}{\partial x_i \partial x_j} R^T ,
\end{equation}
where $R$ represents the rotation matrix between different reference frames. This equation represents the observational equation for the gravity gradients in the frame of gravity field recovery. 

Due to the large amount of observations and the large number of unknowns, it forms a large-scale and over-determined linear equation system for the determination of the gravity field model. The classic Least-Squares (LS) adjustment is applied to solve this linear equation system, which poses a great numerical challenge because of the high computational requirements in terms of both time and memory. The computation of this part is mainly accomplished on the clusters of Leibniz Universit\"at IT Services (LUIS)\footnote{https://www.luis.uni-hannover.de/luis.html}. 

\subsection{Data}
\label{Data}

Three kinds of observations are required for gravity field recovery, including satellite's orbit, attitude and gravity gradients. Orbit data is mainly used to geolocate other observations, while attitude data is necessary for the setup of the rotation matrix between different reference frames. The gravity gradients are the primary observations for the retrieval of the SH coefficients. 

\subsubsection{Synthesis of noise-free data}
\label{Noise_free_obs}

Since the gravity field signal attenuates quickly with altitude, it is important to have the satellite's orbit as low as possible. For the sake of rigorous comparisons to GOCE, we have chosen 71 days of the GOCE orbit, from 2\textsuperscript{nd} March to 10\textsuperscript{th} May, 2013. During this period, the orbit had an average altitude of 239 km. Note that this altitude is lower than the standard one of GOCE, typically of 259 km. The time series of the orbit coordinates have been interpolated at 1 Hz from the GOCE L2 dynamic orbit data sampled at 0.1 Hz using spline functions. Based on the orbit data, the noise-free gravity gradients are computed using the gravity field model ``Eigen-6c4" \cite{Forste2014} up to SH degree and order of 360. 

The generation of the angular velocity $\bm{\Omega}$ is indispensable for the determination of attitude and rotations. This can be done by two steps: first, the instantaneous angular velocity $\omega_{\rm{LORF}}$ of the Local Orbital Reference Frame (LORF) with respect to the Inertial Reference Frame (IRF) has to be computed from the orbital position and velocity, and then assigned to the component $\omega_y$ so that the main rotation of the instrument frame, i.e., GRF, at the orbital frequency is indeed about the y-axis; second, a residual zero-mean angular velocity $\delta \bm{\Omega} = (\delta \omega_x, \delta \omega_y, \delta \omega_z)^T$ is added so that we finally have
\begin{equation}
	\label{angular_velocity}
	\bm{\Omega} = \left(
	\begin{matrix}
		 \delta \omega_x \\
		\omega_{LORF} + \delta \omega_y \\
		\delta \omega_z 
	\end{matrix}	
	\right) .
\end{equation}

\subsubsection{Synthesis of noisy data}
\label{Noisy_obs}

We assumed a zero-mean, normally distributed error with a standard deviation of 2.6 cm added to the orbit initial coordinates. This precision was typically achieved with GOCE \cite{Bock2014}. 

\begin{figure}[!htb]
	\centering
	\includegraphics[width=0.6\textwidth]{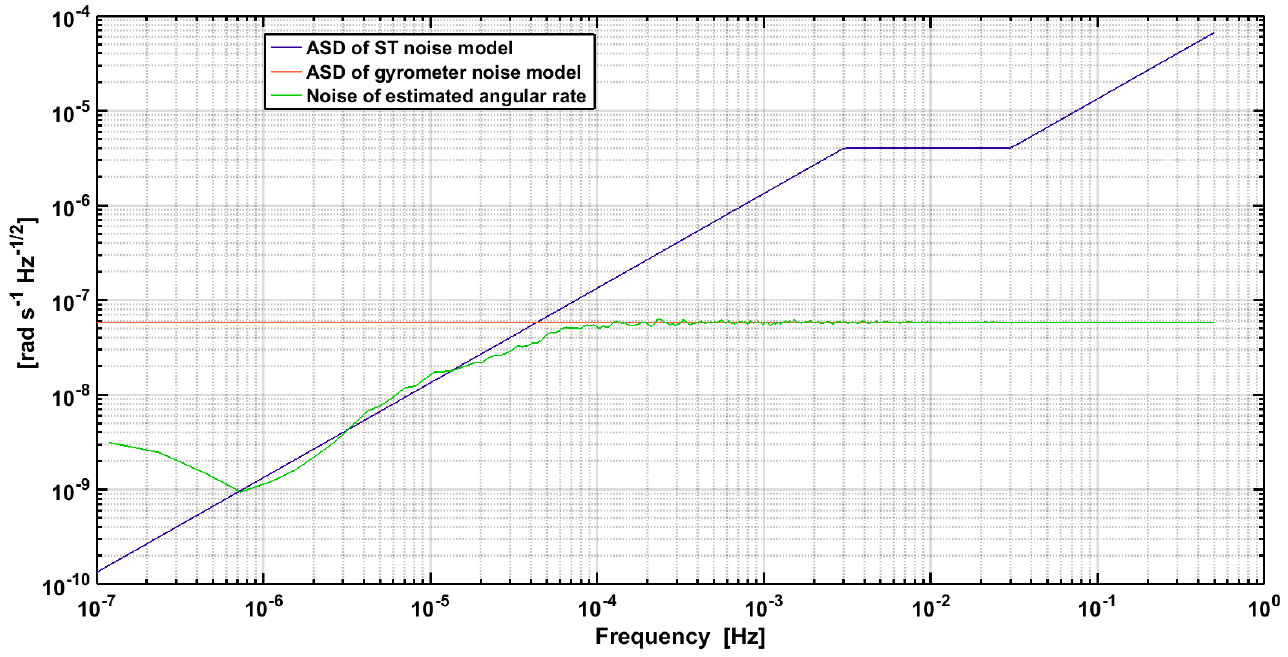}
	\caption{Amplitude spectral density of the noise degrading the estimation of one component of the angular velocity of the instrument frame, i.e., GRF, with respect to the inertial frame.}
	\label{PSD_Angular_Velocity}
\end{figure}

We assumed that three star-trackers and a gyroscope are on board of the satellite for the determination of the angular velocity. Both kinds of sensors give estimates of the three components of the angular velocity, which are then optimally combined in the frequency domain using a Wiener filter like in the processing of the angular velocity of GOCE, see \cite{Stummer2011}. This method requires to have a model of the PSD of the noise of both sensors. Thereafter, the noise models of both sensors were assumed, with their spectral characteristics shown in figure~\ref{PSD_Angular_Velocity}. For more details on this aspect, we refer to reference~\cite{Douch2018}. 
The errors of the gravity gradients are the sum of the gradiometer noise and the error due to the correction of the centrifugal terms. An additive noise with a white behaviour is assumed for the gradiometer. The amplitude of the PSD is assumed as 5 mE/Hz$^{1/2}$. The centrifugal terms are the square of the angular velocities whose noise has been discussed above. 

\subsection{Results}
\label{Results}

Global gravity field models were finally recovered from the synthesized noisy observations. They were resolved up to SH degree and order of 240, including 58 077 unknowns. As pointed out previously, we could have the diagonal gradient components, i.e., $V_{xx}, V_{yy}, V_{zz}$, obtained for the nadir pointing mode with compensation on rotations. We thus derived three component-wise gravity field solutions as well as a combined solution from all three components. To evaluate the performance of these models, both the true errors (the differences between the recovered SH coefficients and the input background model, i.e., Eigen-6c4) and the formal errors (the accompanied standard deviations of the parameters, obtained in the LS adjustment) are analyzed.  

\begin{figure}
	\centering
	\subfigure[$V_{xx}$]{
		\includegraphics[width=0.47\textwidth]{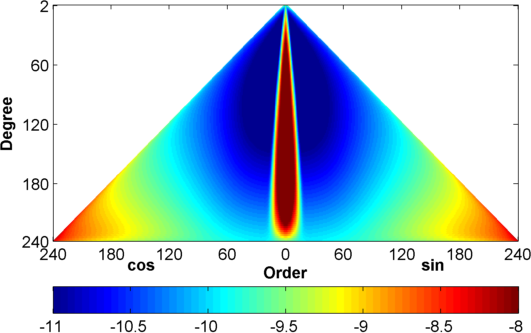}}
		\hfill
	\subfigure[$V_{yy}$]{
		\includegraphics[width=0.47\textwidth]{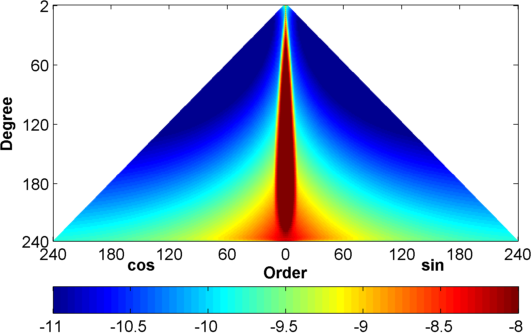}} \\
	\subfigure[$V_{zz}$]{
		\includegraphics[width=0.47\textwidth]{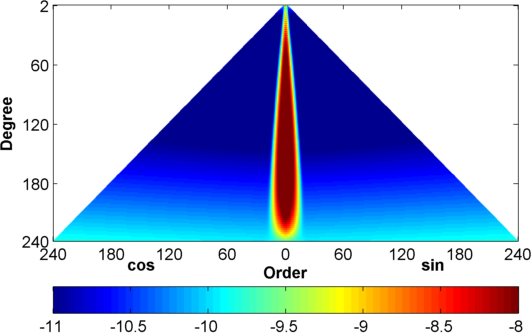}}
		\hfill
	\subfigure[Combined]{
		\includegraphics[width=0.47\textwidth]{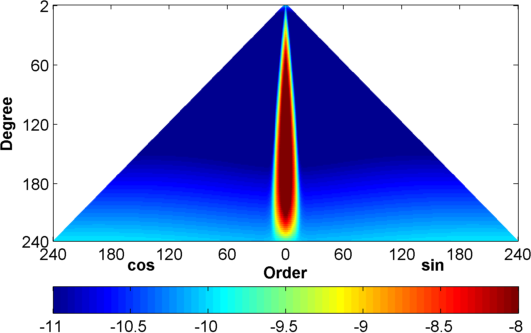}}	
	\caption{Formal errors of the component-wise gravity field solutions, in logarithmic scale.}	
	\label{Formal_errors}
\end{figure}

The formal errors of the component-wise and combined solutions are displayed in figure~\ref{Formal_errors}. It is shown that these gradient components are sensitive to different parts of the gravity field. For instance, $V_{xx}$ is more sensitive to lower order zonal and near-zonal coefficients, i.e., around order zero, but less sensitive to higher order coefficients. In this regard, $V_{yy}$ exactly complements $V_{xx}$ with the inverse sensitivity. It contributes mainly to non-zonal coefficients, i.e., the sectorial coefficients. The reason for the inverse and complementary contribution is related to the orientation of the corresponding gradiometer pairs. Compared to $V_{xx}$ and $V_{yy}$, $V_{zz}$ is sensitive to all orders of the coefficients, according to Laplace's equation, $V_{zz} = - (V_{xx} + V_{yy})$. The contribution of $V_{zz}$ is identical to the combination of $V_{xx}$ and $V_{yy}$. This could be demonstrated indirectly by comparing the $V_{zz}$ component solution with the combined solution, where both show quite similar patterns. Here, we would also like to mention that the zonal and near-zonal coefficients for all solutions are determined in a degraded performance. This is attributed to the GOCE orbit, which leaves polar gaps where no observations are available \cite{Sneeuw1997}. 

The contribution analysis indicates that the one-axis nadir pointing mode shows natural deficits for determining the Earth's gravity field. Only a part of the gravity field signal can be precisely retrieved from the $V_{yy}$ component. However, when the compensation of the rotation was applied in this mode, the other two components $V_{xx}$ and $V_{zz}$ can be obtained with a comparable high accuracy. This three-axis mode can thus integrally capture the gravity field signal. 

\begin{figure}
	\centering
		\includegraphics[width=0.6\textwidth]{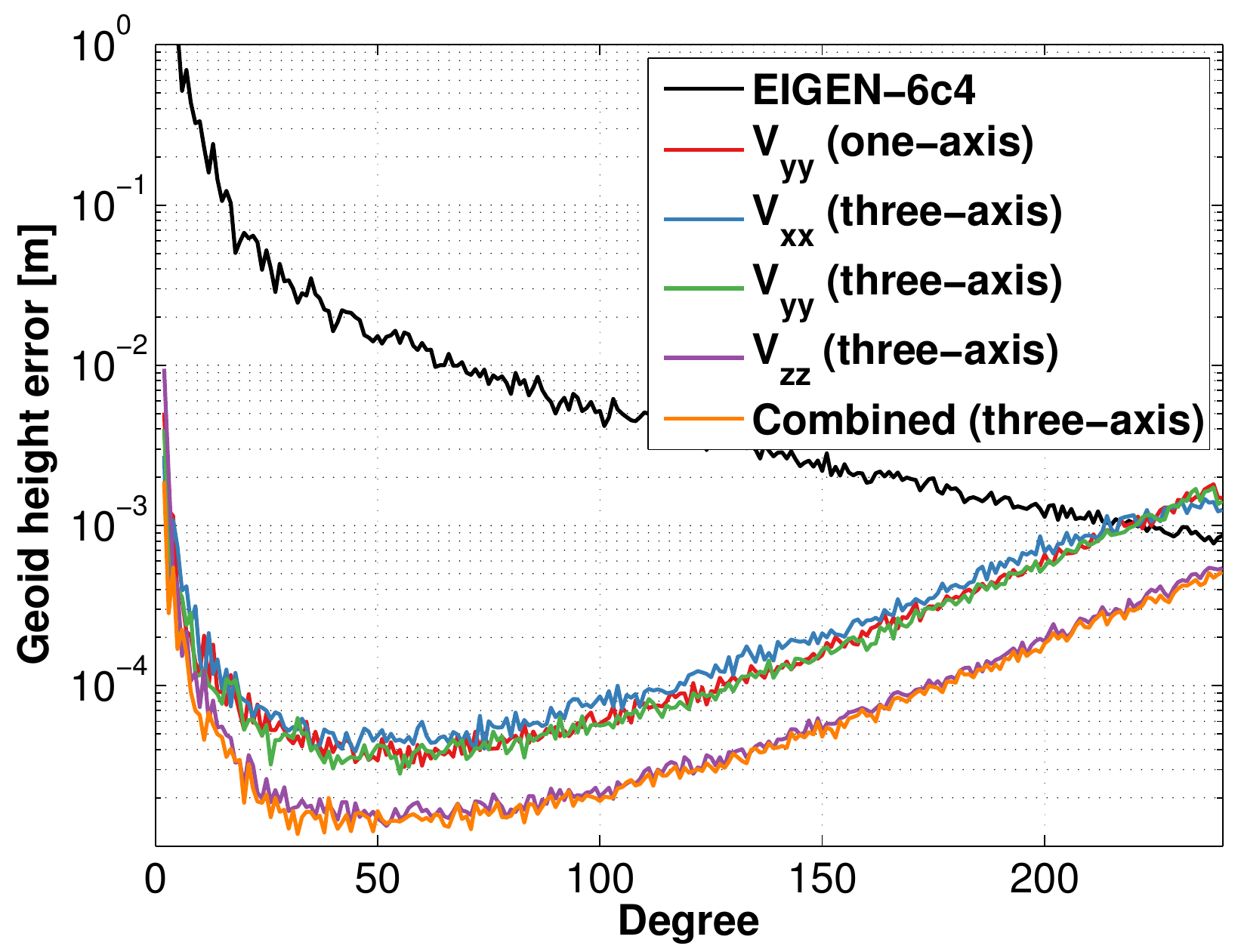}
	\caption{Degree medians of the true errors for the recovered models in three-axis nadir pointing mode, expressed in terms of geoid height. For comparison, the $V_{yy}$ solution in the one-axis mode was plotted as well.}	
	\label{Degree_error}
\end{figure}

The degree medians of the true errors for all recovered models are shown in figure~\ref{Degree_error}. The degree medians that are more robust with respect to the degraded coefficients are used to represent the error amplitudes for each degree. And for a better understanding, the errors are expressed in terms of geoid height. The solutions from $V_{yy}$ in the one-axis and the three-axis mode show the same performance, as the same signal-to-noise ratio (SNR) of the observations has been assumed. Similarly, the error curve of the $V_{xx}$ component is at a comparable level as $V_{yy}$. However, the solution from $V_{zz}$ is about twice better than those of $V_{xx}$ and $V_{yy}$. This is mainly due to the double power of signal, see more discussion in \cite{Douch2018}. To sum up the contribution of all components, the combined solution shows the best performance. It is much better than the component-wise solutions of $V_{xx}$ and $V_{yy}$ but only marginally better than the $V_{zz}$ component solution. The combination of all components by optimal weighting indicates that the combined solution is dominated by the $V_{zz}$ solution. 

\subsection{Discussion}
\label{Discussion}

We analyze now to what extent the CAI gradiometer concept can outperform GOCE when assuming a nominal gradiometer noise of 5 mE/Hz$^{1/2}$. To this end, we extrapolate the error of these models from 71 days to 8 months assuming the error is purely stochastic and reduces as $\sqrt{t}$ where $t$ is the time of integration. To gain an idea of the GOCE solution error, we compute a gravity field model based on the three diagonal gravity gradients of the whole GOCE mission period (November 2009 -- October 2013, about 47 months). Since the comparison concerns only the contribution of the gradiometer, we have not taken into account the GOCE high-low satellite-to-satellite tracking data that is mainly responsible for the recovery of the low-degree gravity field coefficients. 

\begin{figure}[!htbp]
	\centering
	\includegraphics[width=0.6\textwidth]{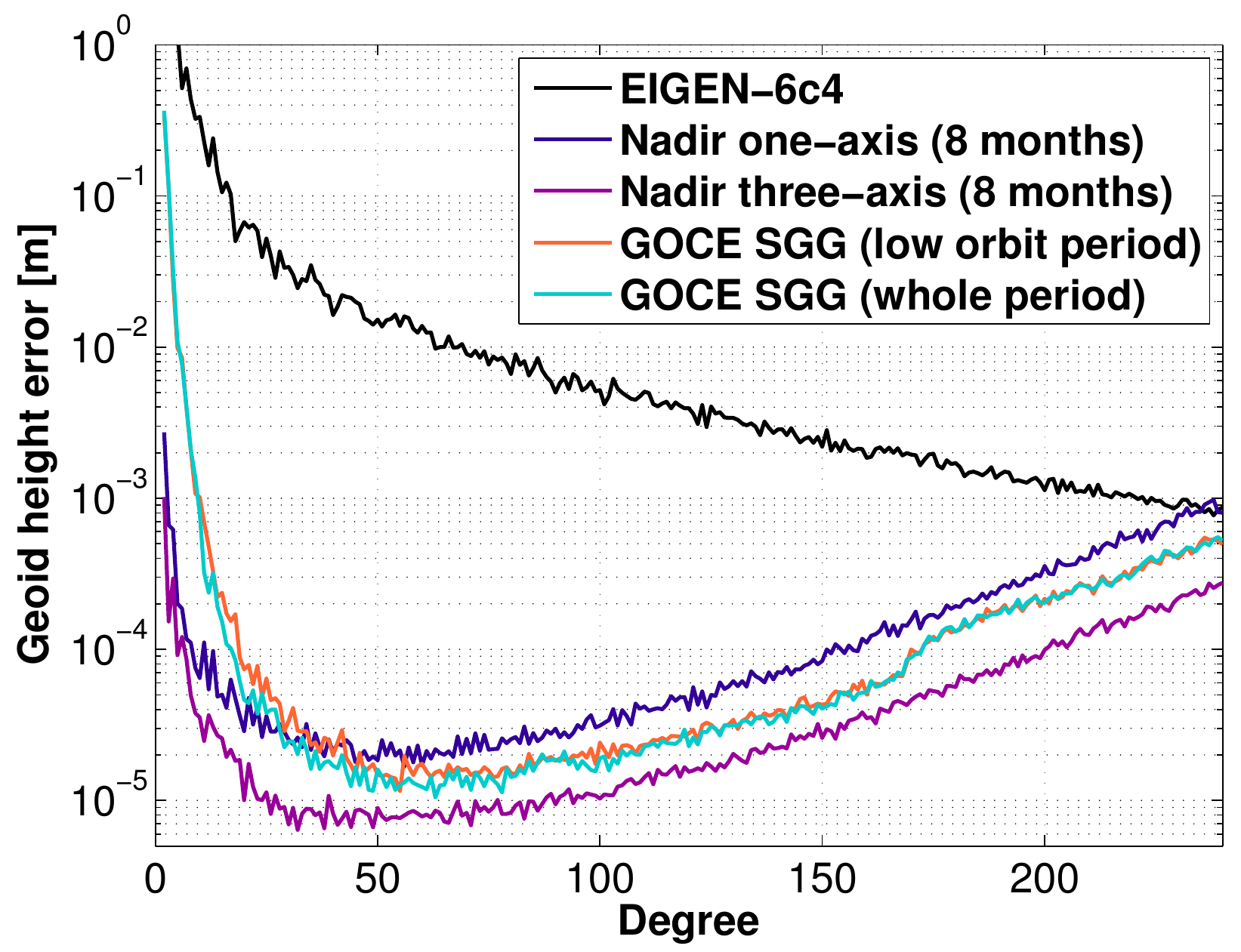}
	\caption{Degree medians of the true errors for gravity field solutions that are extrapolated to 8 months for the CAI gradiometer as well as the errors for the GOCE SGG (Satellite Gravity Gradient) solutions. Two GOCE SGG solutions are included here. One is based on the gravity gradients of the whole mission period, while the other one corresponds to the period where the satellite was in lower orbits.} 
	\label{Compare_GOCE}
\end{figure}

The 8-month solution for the three-axis nadir pointing mode is better than the GOCE solution for the whole mission period, as shown in figure~\ref{Compare_GOCE}. We can thus conclude that an 8-month mission at an altitude of 239 km and using a 3-axis CAI gradiometer in the nadir pointing mode with a nominal white noise of 5 mE/Hz$^{1/2}$ would outperform the full GOCE mission and eventually yields a more precise gravity field model. Nonetheless, in this comparison, it should be kept in mind that the GOCE satellite was most of its lifetime at a altitude higher than 239 km, about 3 months at 239 km and 5 months lower than 239 km. This fact shows that it is technically possible to fly a satellite at an altitude as low as 239 km for a duration of 8 months. 

For a fair comparison, we have also plotted the GOCE solution for the 8 months where the satellite was at an altitude equal or lower than 239 km. Again, the 8-month CAI gradiometer in the three-axis nadir mode at a constant altitude yields a better solution despite the fact that the 8-month GOCE solution is partly based on gravity gradients measured at a lower altitude.

\section{Conclusion} 

We have carried out a detailed design of a 3D gradiometer based on atom interferometry for space geodesy. We have performed a detailed analysis and modelling of the atomic signals and of the constraints on relevant parameters (i.e. atomic source, interferometer geometry and attitude control of the satellite). The implementation of cancellation methods for the large rotation rate at the orbital frequency when operating Nadir, and for the gravity gradient, allows for reducing dephasing and systematic effects and for extracting the signal with maximal sensitivity. With an expected sensitivity of 5 mE/Hz$^{1/2}$ (PSD), we show a two-fold improvement on the gravity field recovery for degrees above 50, and significantly better for lower orders, when comparing an 8-month solution at an altitude of 239 km with the model obtained from GOCE data over its whole mission duration. 

The determination of the optimal gain requires a realistic mission scenario, which remains to be investigated. As inputs, such mission-oriented study would use the constraints which we have determined for the attitude control and the overall size, weight and power (SWaP) budget of the total instrument. This budget has been established considering existing and available technology, and certainly needs to be reduced to end up with a more reasonable load. Possible modifications to the design, such as sharing subsystems between the instruments, would certainly help, but it is clear that a number of specific technological and engineering efforts are also required, in particular directed towards the optimization of the power consumption. This challenging task motivates on-going and future research and development activities. This concerns not only the technological efforts mentioned above, to improve for instance the generation of BEC sources on atom chips or the compactness and power consumption of fiber-based laser systems, but also the validation of the instrument concept. Indeed, if the key scientific methods, such as bloch-lattice transport, double Raman diffraction or interleaved measurements, have for most of them been demonstrated individually, demonstration activities combining several, and in the end all, of them in a single setup in a representative environment, need to be pushed. This calls for carrying prototyping activities, such as developing an elegant breadboard model of the sensor and characterizing it in a relevant environment. A thorough assessment of the performances of such a prototype will establish gradiometers based on cold atom interferometry as appealing sensors for future gravity missions aiming at improving our knowledge of the Earth's gravity field.

\section*{Acknowledgments} 

\addcontentsline{toc}{section}{Acknowledgments} 

This work has been carried out in the context of the ``Study of a Cold Atom Interferometer Gravity Gradiometer Sensor and Mission Concepts", supported by the European Space Agency through Contract No. 4000112677/14/NL/MP.
The authors affiliated to the IQ acknowledge financial support by ``Nieders\"achsisches Vorab" through the ``Quantum- and Nano- Metrology (QUANOMET)" initiative within the project QT3, and by the German Space Agency DLR with funds provided by the Federal Ministry of Economics and Technology (BMWi) under the grant numbers 50 WP 1431 and 1700.


\pagebreak
\bibliographystyle{unsrt}
\bibliography{CAI_2}

\end{document}